%% file: JSAC2014_InterferenceCoordination_ver14.tex
\documentclass[draftclsnofoot,onecolumn,12pt]{IEEEtran}
\usepackage{amsthm,amssymb,graphicx,multirow,amsmath,color,amsfonts}%,ulem}
\usepackage[update,prepend]{epstopdf}
\usepackage[noadjust]{cite}
\usepackage[latin1]{inputenc}
\usepackage{tikz}
\usetikzlibrary{arrows,calc}		% Optional ticks libraries
\usepackage{bbm} 				% for \mathbbm{1}
\usepackage{pdfpages}
\usepackage{flushend,subfigure}
\usepackage{tabulary}
\usepackage{multirow}
\usepackage{comment}

\usepackage[margin=1in]{geometry}

\include{notation}
\include{notation_HD}

\input{macros}

\allowdisplaybreaks

\begin{document}
\graphicspath{{./PaperFigures/}}
\title{Massive-MIMO Meets HetNet: Interference Coordination Through Spatial Blanking
%Spatial Blanking and Inter-Tier Coordination in Massive-MIMO Heterogeneous Cellular Networks
%Inter-Tier Interference Coordination in Massive-MIMO Heterogeneous Cellular Networks
}
\author{ Ansuman Adhikary, Harpreet S. Dhillon, and Giuseppe Caire
%	Harpreet S. Dhillon,~\IEEEmembership{Member,~IEEE,}
        %ABC,~\IEEEmembership{Fellow,~IEEE,}
        %and~XYZ,~\IEEEmembership{Life~Fellow,~IEEE}% <-this % stops a space
%	\thanks{H. S. Dhillon is with the Communication Sciences Institute (CSI), Department of Electrical Engineering, University of Southern California, Los Angeles, CA (email: hdhillon@usc.edu).}%
	\thanks{The authors are with the Communication Sciences Institute (CSI), Department of Electrical Engineering, University of Southern California, Los Angeles, CA (email: \{adhikary; hdhillon; caire\}@usc.edu).} 
	\thanks{This paper is submitted in part to IEEE Globecom, Workshop on Heterogeneous and Small Cell Networks (HetSNets), Austin, TX, Dec. 2014~\cite{AdhDhiC2014}. \hfill
	Manuscript last updated: \today.}%
	
}

%\markboth{IEEE Transactions on Communications}%
%{Submitted paper}

\maketitle

\begin{abstract}
In this paper, we study the downlink performance of a heterogeneous cellular network (HetNet) where both macro and small cells share the same spectrum and hence interfere with each other. We assume that the users are concentrated at certain areas in the cell, i.e., they form {\em hotspots}. While some of the hotspots are assumed to have a small cell in their vicinity, the others are directly served by the macrocell. Due to a relatively small area of each hotspot, the users lying in a particular hotspot appear to be {\em almost} co-located to the macrocells, which are typically deployed at some elevation. Assuming large number of antennas at the macrocell, we exploit this directionality in the channel vectors to obtain {\em spatial blanking}, i.e., concentrating transmission energy only in certain directions while creating transmission opportunities for the small cells lying in the other directions. In addition to this inherent interference suppression, we also develop three low-complexity interference coordination strategies: (i) turn off small cells based on the amount of cross-tier interference they receive or cause to the scheduled macrocell hotspots, (ii) schedule hotspots such that treating interference as noise is approximately {\em optimal} for the resulting Gaussian interference channel, and
%select the set of active transmitters such that treating interference as noise is {\em optimal} for all the links, and
(iii) offload some of the macrocell hotspots to nearby small cells in order to improve throughput fairness across all hotspots. For all these schemes, we study the relative merits and demerits of uniform deployment of small cells vs. deploying more small cells towards the cell center or the cell edge. %Overall, the idea of ``spatial blanking'' is significantly more efficient than isotropic slot blanking being considered for eICIC in 3GPP.
\end{abstract}

\begin{IEEEkeywords}
Massive-MIMO, heterogeneous cellular network, hotspots, interference coordination, spatial blanking.
\end{IEEEkeywords}

\section{Introduction}
Since the introduction of advanced communication devices, such as smartphones, tablets and laptops, the demand for mobile data traffic is almost getting doubled every year and the trend is expected to continue for at least a few more years~\cite{CisM2012}. In typical dense urban scenarios, a large proportion of data traffic is generated by highly concentrated groups of users, e.g., at the coffee shops or the airports, which are often termed as {\em hotspots}. A possible solution to handle this data demand is by deploying a large number of low power base stations, called {\em small cells}, especially closer to the areas of high user density (hotspots). If these small cells are operated in the same spectrum as the macrocells (typically the case in cellular) both the inter-tier and intra-tier interference threaten the gains achieved by densification. Mitigating this interference by {\em spatial blanking} and low complexity inter-tier coordination is the main focus of this paper.

\subsection{Related Work}
Interference coordination in HetNets has attracted a lot of attention both in standards bodies, such as 3GPP, as well as in academic research, e.g., see \cite{eicicperez,ghosh2012heterogeneous} and the references therein. The typical approach for mitigating this interference involves orthogonalizing the time-frequency resources allocated to the macrocells and small cells. This is also the main objective of 3GPP's enhanced Inter-Cell Interference Coordination (eICIC), which allows orthogonalization over both time and frequency. Orthogonalization over frequency can be achieved through {\em fractional frequency reuse}, where the users in the interior of all the cells are scheduled on the same frequency resources, whereas the users at the cell edge of the neighboring cells are scheduled on orthogonal resources to mitigate interference~\cite{BouPanJ2009}. Orthogonalization over time is achieved by introducing {\em almost-blank subframes} where the idea is to blank (turn off) some sub-frames of the macrocell so as to reduce the inter-tier interference caused to the small cells~\cite{eicicperez}. Note that the interference coordination can also be achieved through distance-based power control, as demonstrated in the case of ``cognitive'' small cells in~\cite{adhikary2011cognitive}.

In the literature, there are two main directions taken for the analysis of interference coordination in HetNets. The first one focuses mainly on the spatial aspects (geometry) of the network, where both macrocells and small cells are modeled as independent Poisson Point Processes (PPPs) and tools from stochastic geometry are used to derive easy to use expressions for key performance metrics, such as coverage and rate~\cite{DhiGanJ2012}. Since these results are {\em averaged} over network realizations, these are ideal to understand the ``macroscopic'' performance trends.
The performance of fractional frequency reuse in single antenna HetNets is studied using these tools in \cite{NovGanJ2012}. The performance of almost-blank subframes is also studied using these tools in~\cite{SinAndJ2014,CieWanJ2013}, where the general idea is to derive a tractable expression for the performance metric, e.g., downlink rate distribution, in terms of the fraction of the sub-frame that is blanked along with other parameters, such as transmit power, and the macrocell and small cell densities. This fraction can then be tuned to optimize the given performance metric. While the performance of multi-antenna HetNets using these tools is under investigation~\cite{DhiKouJ2013,GupDhiJ2013}, the analysis of interference coordination for this case does not appear in the literature. Moreover, active interference coordination using antenna beamforming techniques introduces statistical dependency among macro and small cells,  which in turns makes the use of standard stochastic geometry tools very difficult in general. The second direction of analysis is more suited for multi-antenna HetNets, especially when the number of antennas is large (commonly referred to as ``massive-MIMO''). The main idea is to use random matrix theory results to reduce channel gains to deterministic constants, which simplifies the analysis significantly~\cite{marzetta2010noncooperative,wagner2012large}. This forms the foundation of massive-MIMO analysis, which is also a key element of this paper. %Some of the related works in this direction are summarized next.

%The typical approach for mitigating this interference, e.g., by enhanced Inter-Cell Interference Coordination (eICIC) in 3GPP, involves orthogonalizing the time frequency resources allocated to the macro and small cells. When both the macrocell as well as the small cells share the same spectrum, ``cognitive'' small cells, proposed in \cite{adhikary2011cognitive}, can provide very high area spectral efficiency (bit/s/Hz/km$^2$) in both tiers with a simple distance based power control and moderate number of antennas at the small cells.

Having a large number of antennas at the macrocells provides an interesting alternative to interference coordination without the need for orthogonalizing resources over time or frequency. Since the macrocell is typically located at an elevated position (e.g., tower-mounted, or deployed on a building roof), it ``sees'' both its own users as well as the small cells under a relatively narrow angular spread. This gives rise to highly directional channel vectors, which can be modeled as Gaussian random vectors
with a small number of dominant eigenmodes (eigenvectors of their covariance matrix). The macrocell can exploit this directionality by using Joint Spatial Division and Multiplexing (JSDM), proposed in \cite{adhikary2012joint}, in order to simultaneously provide spatial multiplexing to its own users as well as mitigate the cross-tier interference to the small cells. This can be achieved explicitly by nulling certain spatial directions, i.e., by transmitting in the orthogonal complement of the dominant eigenmodes of the channel vectors from the macrocell to a subset of selected small cells or implicitly by serving users that are not in the direction of the small cells. We call this approach {\em spatial blanking}, in analogy with the almost-blank subframe approach of eICIC. Note that JSDM has been considered for inter-tier coordination in the context of cognitive small cells and reverse time division duplex (TDD) architecture in \cite{adhikary2014massive}.

\subsection{Contributions}

{\em Realistic model.} We propose a realistic HetNet setup with two key features: (i) users are concentrated at certain areas in the cell, thereby forming hotspots, and (ii) small cells are deployed in the vicinity of some hotspots, as is typically the case in current capacity-driven deployments. The rest of the hotspots are directly served by the macrocell. Assuming that the hotspot sizes are much smaller than the macrocell radius, the first feature facilitates {\em spatial blanking}, under which a massive-MIMO macrocell can focus its energy in the direction of hotspots that it serves, while allowing the simultaneous transmission of small cells located in the other directions. The second feature allows low-complexity cell coordination strategies by simplifying the cell selection procedure significantly.

{\em Inter-tier coordination strategies.} While spatial blanking provides an implicit interference mitigation, the {\em random} geometry of the hotspots may still result in a high inter-tier interference. For instance, a hotspot served by a small cell may lie in the same direction as a hotspot being served by the macrocell, thereby experiencing a strong inter-tier interference. To mitigate interference in such cases, we develop three low-complexity interference coordination strategies. In the first strategy, a small cell shuts down its transmission based on the amount of cross-tier interference it receives or causes to the scheduled macrocell hotspots. The second strategy selects the set of active small cells such that treating interference at all the scheduled hotspots is optimal. On the other hand, the third strategy ``offloads'' some of the macrocell hotspots to small cells, thereby increasing fairness in the rates of all the hotspots.

{\em System design guidelines.} An important consequence of spatial blanking and directional channels is that it is no longer {\em a priori} intuitive whether deploying more small cells at the cell edge provides the best performance, as is the case in single antenna HetNets. We therefore, compare the merits and demerits of the above coordination strategies in three deployment scenarios: (i) small cells uniformly distributed over a macrocell, (ii) more small cells towards the center, and (iii) more small cells at the cell edge. While the cell edge deployment indeed turns out to be better in most cases, the gains are much higher when the macrocell traffic is offloaded to small cells, which increases fairness in the rates across hotspots. More aggressive offloading of macrocell hotspots increases their rate while reducing the rate of small cells, which may not eventually affect the system performance because of the limited backhaul capacity of the small cells. Please refer to~\cite{SinDhiJ2013,AndSinJ2014} for more insights on traffic offloading and load balancing in HetNets.

%whether deploying more small cells at the cell edge provides the best performance,  as in the case of ...

%%%%%%%%%%%%%%%%%%%%%%%%%%%%%
\section{System Model}  \label{sec:system-model}

We consider a HetNet formed by macrocell base stations coexisting in the same coverage area
and channel frequency band with small cell base stations. Both macrocells and small cells are equipped with multiple antennas and
serve multiple users in each transmission time-frequency slot (denoted hereafter as a {\em transmission resource block}).
In particular, we denote by $M$ and $L$ the number of antennas at the macrocells and at the small cells, respectively, and assume
the regime of {\em massive-MIMO} \cite{marzetta2010noncooperative}, \cite{huh2012achieving}, for which $M \gg L \gg 1$, and the number of users simultaneously served by each base station is significantly less than the corresponding number of transmit antennas. Note that this setup where massive-MIMO macrocells coexist with multi-antenna small cells is well motivated in the context of 5G cellular networks \cite{JunManJ2014}.

%%%%%%%%%%%%%%%%%%%%%%%%%%%%%%%%%%%%
\subsection{Spatial Setup}

We assume a single-cell scenario, with a macrocell located at the center of a disk of radius $R_{\rm mc}$,
where $R_{\rm mc}$ denotes the macrocell coverage radius.\footnote{The extension of this work to a multi-cell setting without any explicit coordination across macrocells is straightforward.}
We focus on a non-uniform user distribution. In particular, we assume that the users are clustered into several high density {\em hotspots},
referred hereafter as {\em user groups}. We assume $N_{\rm u}$ user groups uniformly and independently distributed in the macrocell area.
Each group is concentrated over an area much smaller than the macrocell disk. Furthermore, we assume that the scattering ``landscape''
for the users in the same group is the same, while the users are separated by several wavelengths. For example,
consider a cluster of users located outdoor, such as a bus stop. They are physically separated by at least 1m, spanning at least six wavelengths at a
carrier frequency of 2GHz. However, the scattering landscape that determines the angular distribution of the propagation between
such users and the macrocell, and the distance-dependent pathloss, are virtually identical for all such users.
A similar consideration can be made for indoor users clustered in small environments, such as a coffee shop.
Hence, for the sake of mathematical simplicity, we shall treat the user groups as ``co-located''. This implies that the user channel vectors
from users in the same group to the macrocell antenna array are mutually independent (due to the several wavelength separation between the users) but
are identically distributed with the same covariance matrix (due to the fact that the scattering landscape is the same for all such users) \cite{adhikary2012joint}, \cite{lee1973effects}, \cite{molisch2010wireless}. We defer more details to Sections~\ref{subsec:macro} and~\ref{subsec:femto}.

In order to capture the fact that the position of the user groups is evolving in time, and therefore it is not possible to cover each group with a dedicated
small cell, we let $N_{\rm f} \leq N_{\rm u}$ denote the number of small cells in the system, each of which covers a user group.
The set of such user groups is denoted by $\Sc$. For convenience,  the small cell is assumed to be located at the center of its user group,
reflecting a small cell deployed at known and persistent hotspots such as airport lounges or coffee shops.
The remaining $N_{\rm u} - N_{\rm f}$  user groups, denoted by the set $\Mc$,
can be served either by the macrocell or by some neighboring small cell through {\em offloading}, the exact details of which will
appear in Section \ref{sec:inter-tier-ic}.

\begin{figure}[!h]
  \centering
  \includegraphics[width=0.32 \textwidth]{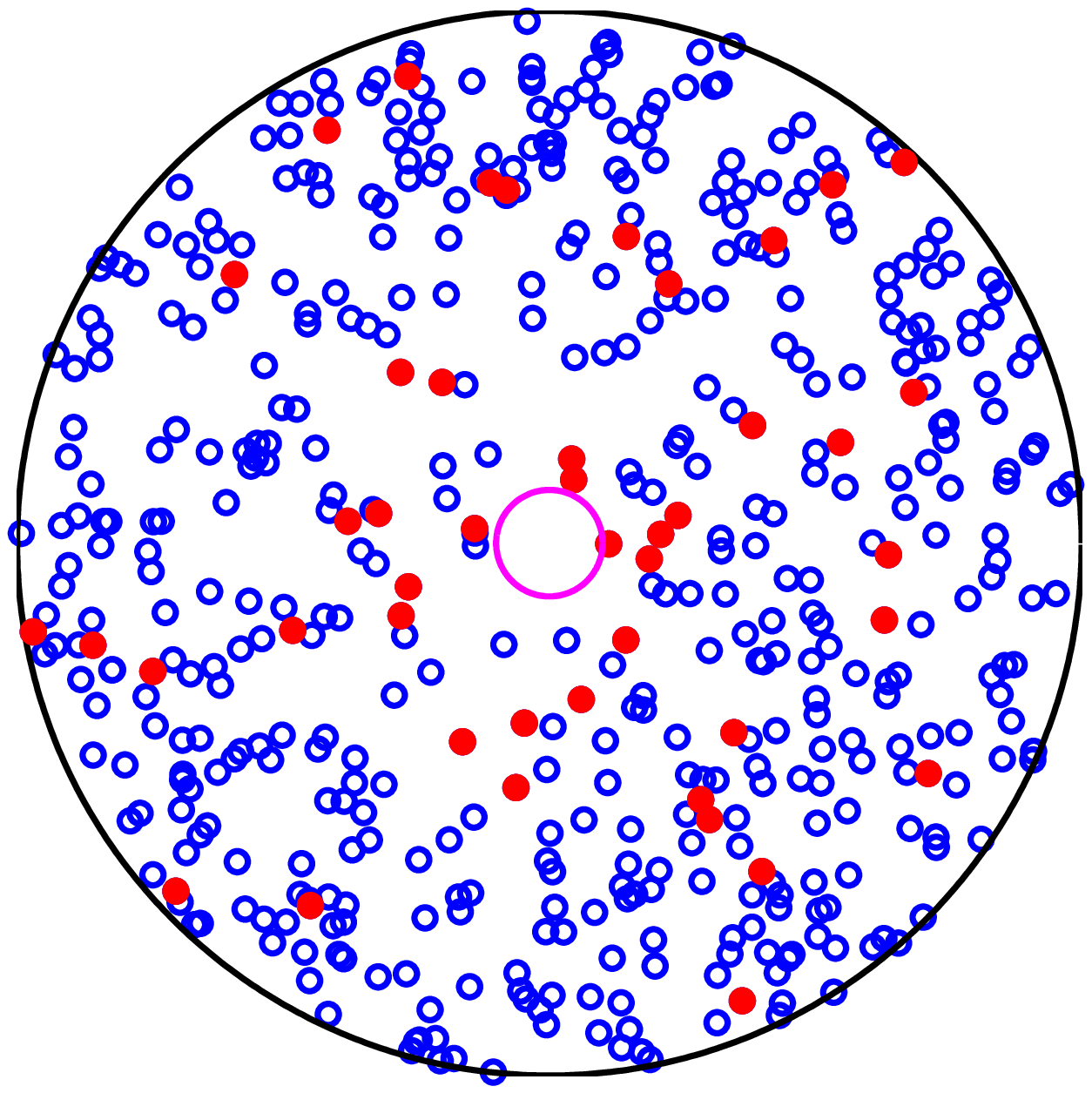}
  \includegraphics[width=0.32 \textwidth]{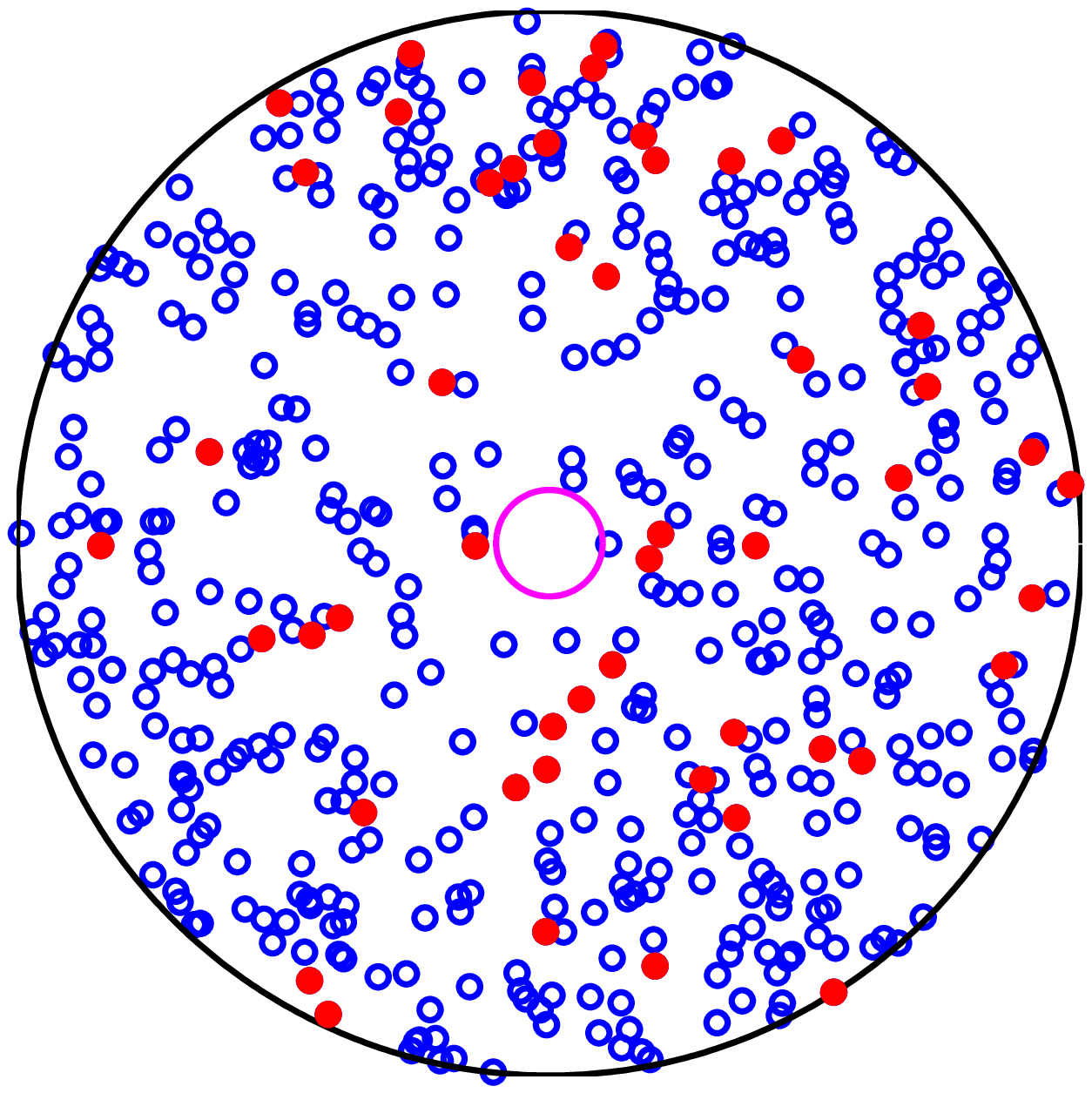}
  \includegraphics[width=0.32 \textwidth]{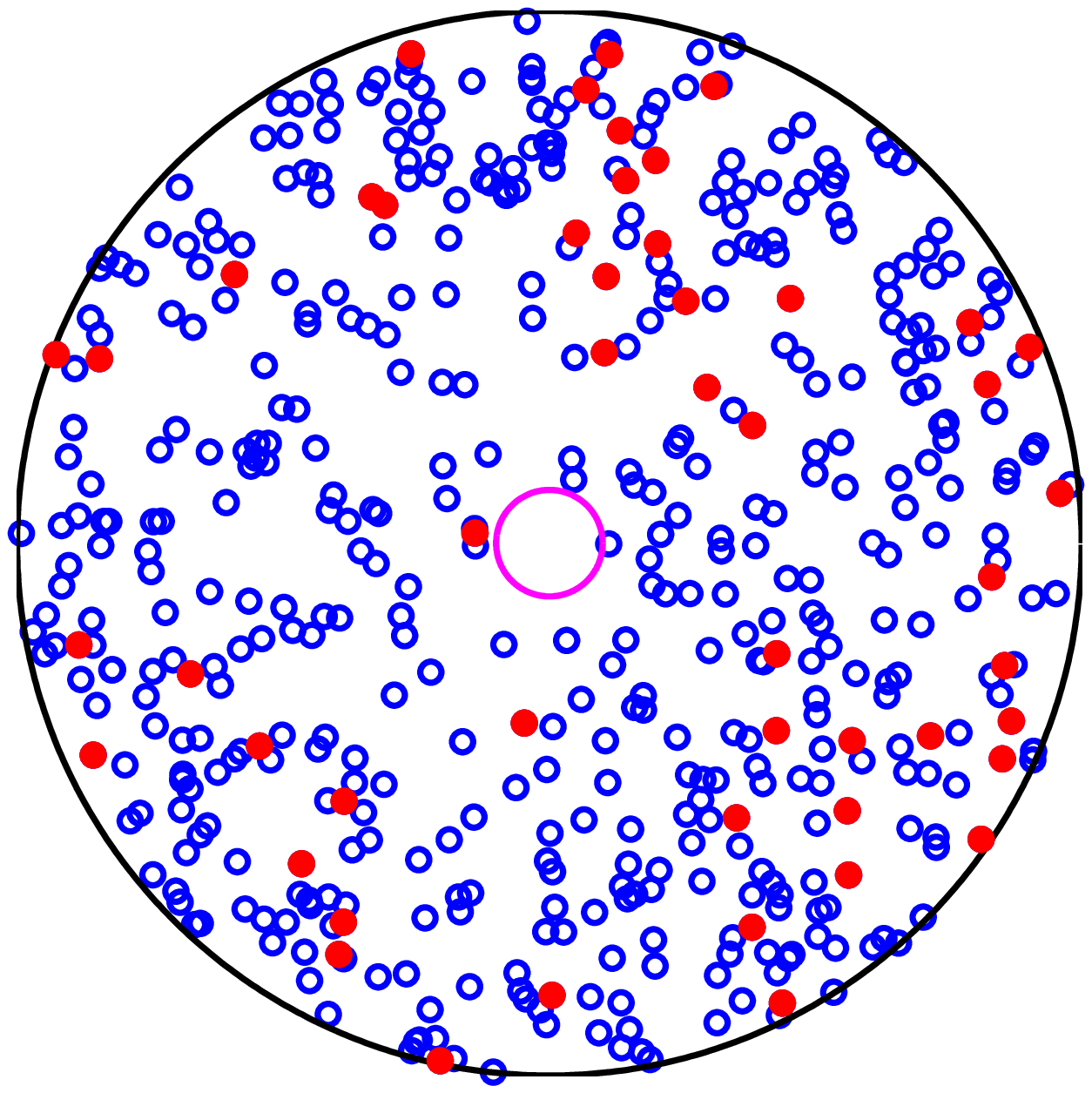}
  \caption{The hollow ``blue'' circles denote the user groups that do not contain a small cell, and the filled ``red'' circles denote the user groups containing a small cell.
  {\em (first)} Cell-interior deployment. {\em (second)} Uniform deployment. {\em (third)} Cell-edge deployment.
  The macrocell is located at the center of the ``magenta'' circle, which denotes an exclusion radius for the macrocell.}\label{fig:layout}
\end{figure}

A natural question that arises now is: given that the user groups are uniformly distributed in the macrocell disk,
how should the set $\Sc$ be chosen? For instance, is it better to deploy the small cells also
uniformly across the whole cell or more concentrated on the cell edge rather than towards the cell center?
The answer to this question is not straightforward {\em a priori} due to the possibility of ``spatial blanking''. In order to address this question we consider three options for small cell deployment defined as follows:

\begin{definition}[Small cell deployment scenarios] \label{def:smallcellscenarios}
We define the following three small cell deployment scenarios to be used in this paper. (i) {\em Uniform deployment:}
small cells are randomly assigned to $N_{\rm f} < N_{\rm u}$ user groups; with uniform probability over all possible
${N_{\rm u} \choose N_{\rm f}}$ assignments.  (ii) {\em Cell-interior deployment:} small cells are randomly assigned to $N_{\rm f} < N_{\rm u}$ user groups
with a distribution that concentrates them more towards the center of the macrocell disk. This is obtained by randomly selecting $N_{\rm f}$ user groups
such that the probability distribution function of the distance $R$ of the selected groups from the disk center is $F_R(r) = \frac{r}{R_{\rm mc}}$.
(iii) {\em Cell-edge deployment:} small cells are randomly assigned to $N_{\rm f} < N_{\rm u}$ user groups
such that the probability distribution function of the distance $R$ of the selected groups from the disk center is $F_R(r) = \frac{r^3}{R_{\rm mc}^3}$.
\end{definition}
%\begin{itemize}
%\item {\em Uniform.} This corresponds to a uniform deployment of small cells over the given macrocell. This is done by randomly assigning small cells to $N_{\rm f} < N_{\rm u}$ user groups.
%\item {\em Center.} This corresponds to a non-uniform deployment of small cells, which are concentrated towards the center of the macrocell and their density progressively decreases as we move towards the cell edge. In order to model such a deployment, we pick $N_{\rm f}$ user groups such that the probability distribution function of the distance $R$ of the small cells from the macrocell center is $F_R(r) = \frac{r}{R_{\rm mc}}$.
%\item {\em Edge.} This corresponds to the other extreme, where the small cells are more concentrated near the edge of the macrocell. In this case, $N_{\rm f}$ user groups are assigned a small cell such that the probability distribution function of the distance $R$ between the user group and the macrocell center is $F_R(r) = \frac{r^3}{R_{\rm mc}^3}$.
%\end{itemize}
A snapshot realization of the three deployment scenarios is shown in Fig.~\ref{fig:layout}.

\subsection{Macrocell Subsystem} \label{subsec:macro}
\setcounter{paragraph}{0}

We assume each user group is surrounded by a circular ring of scatterers of radius $R_{\rm u} \ll R_{\rm mc}$,
giving rise to the one-ring channel model between the macrocell and each user group.\footnote{Interested readers should refer to \cite{adhikary2013joint2} for the discussion on the validity of this statement,
and \cite{adhikary2013joint} for efficient user grouping algorithms.}
As discussed before,  the channel vectors from the macrocell antenna array to the users in the same group
are mutually independent but identically distributed, with the {\em same} channel covariance matrix.
We assume that, at every transmission resource block, the macrocell schedules $G \leq N_{\rm u}$
user groups, where $G$ is a design parameter that governs the throughput tradeoff region. For each of the scheduled user groups,
a subset of the users is served by spatial multiplexing. The number of served users in each selected group $g$, denoted by $S_g$,
depends on the rank of the user group channel covariance matrix and can be optimally selected as discussed extensively in
\cite{adhikary2012joint}.
We further assume that the set of $G$ user groups selected by the macrocell to be served simultaneously are widely separated
in their angular scattering components, so that the dominant eigenspaces of the corresponding channel covariance matrices
are linearly independent.

The instantaneous channel between a user $k$ in group $g$ (denoted by $g_k$) and the macrocell, over any given transmission resource block,
is an $M \times 1$ Gaussian random vector denoted by $\hv_{g_k,0}$.  Using the Karhunen-Loeve representation, we can write
\begin{equation} \label{hgk}
\hv_{g_k,0} = \Um_g \Lambdam_g^{1/2} \wv_{g_k},
\end{equation}
where $\Rm_g = \Um_g \Lambdam_g \Um_g^\herm$ is the channel covariance matrix of rank $r_g$, common to all users in group $g$,
$\Um_g$ is the tall unitary matrix of eigenvectors, of dimension $M \times r_g$, $\Lambdam_g$ is the $r_g \times r_g$
diagonal positive definite matrix of covariance eigenvalues (Karhunen-Loeve coefficients). The $r_g \times 1$ random vector $\wv_{g_k} \sim \Cc\Nc(0,\Id_{r_g})$ is independent for different users and corresponds to the
randomness due to the small-scale multipath fading components.\footnote{The typical duration over which the channel covariances change is several orders of magnitude larger than the dynamics of small-scale fading. Therefore, for mathematical convenience, we assume $\Rm_g$ to be fixed in time and
consider average rates (i.e., ergodic rates) with respect to the small-scale fading components. Notice that under the classical Wide-Sense Stationary Uncorrelated Scattering channel model \cite{molisch2010wireless}, the channel process is wide-sense stationary and therefore its second-order statistics are constant in time, as we assume here. This assumption is valid ``locally'' when observing the system on the time-scale of a few tens of seconds.  In practice, the channel covariance matrices must be adaptively learned and tracked in order to follow the non-stationary time-varying effects in the network (e.g., due to user mobility).}

Following \cite{adhikary2012joint}, we consider the one-ring scattering model in order to determine $\Rm_g$.
Namely, for a user group located at an angle of arrival $\theta_g$ and having angular spread $\Delta_g$, we have
$\Rm_g = \Rm(\theta_g,\Delta_g)$ where, assuming a uniform linear array at the macrocell, the element $(m,n)$ of $\Rm(\theta_g,\Delta_g)$ is given by
    \begin{equation}  \label{one-ring}
    \left[\Rm(\theta_g,\Delta_g)\right]_{m,n} = \frac{a_{g,0}}{2 \Delta_g} \int_{\theta_g-\Delta_g}^{\theta_g+\Delta_g} e^{-j \pi (m - n) \sin(\alpha)} {\rm d}\alpha
    \end{equation}
and $a_{g,0}$ represents the path loss due to the propagation environment, which is given in (\ref{eqn:path-loss}).

The total macrocell transmit power is denoted by $P_0$.
For analytical simplicity we consider equal power allocation, such that all the macrocell downlink data streams are transmitted with the same power
$\frac{P_0}{S}$, where $S \leq \sum_{g=1}^G S_g$ is the total number of downlink streams, i.e., the total number of served users across all groups.

\subsection{Small Cell Subsystem} \label{subsec:femto} \setcounter{paragraph}{0}

Small cells serve multiple users by spatial multiplexing.
We assume that in each transmission resource block, a small cell  serves $\bar{S}_f$ users in the group.\footnote{We assume that the user groups are fully loaded, i.e., they contain a sufficient number of users (much larger than $\bar{S}_f$).}
We also assume the presence of coordination between the small cells and the macrocell, in order to implement some form of {\em inter-tier interference coordination} as discussed in Section \ref{sec:inter-tier-ic}.
We assume that all the active small cells transmit at their peak power $P_1$.
The presence of multiple users gives rise to intra-cell interference, which is handled by
zero forcing beamforming. In the regime of massive-MIMO, it is well known that {\em user selection} \cite{yoo2006optimality}, \cite{yoo2007multi}, \cite{dimic2005downlink} yields negligible gains at the cost of a high channel state information overhead. Therefore, we assume that each small cell simply schedules a random set of
$\bar{S}_f$ users in its group with uniform probability, achieving proportional fairness (which in this case coincides with
equal air-time) across all its users.

Since the small cells are typically deployed at low elevation, the scattering geometry of the channels between users and the small cell array can be considered isotropic. Hence, the channel between a user $k$ in group $g$ and a small cell in group $f$ is modeled as a $L \times 1$ vector $\hv_{g_k,f}$ with i.i.d. entries
$\sim \Cc\Nc(0, a(g,f))$, where $a(g,f)$ is the distance-dependent path-loss coefficient between the users in group $g$ and the small cell co-located with group $f$, and is given by
\begin{equation} \label{eqn:path-loss}
a(g,f) = \frac{w^{n_w(g,f)}}{1 + \left(\frac{d(g,f)}{d_0}\right)^\alpha}
\end{equation}
where $d(g,f)$ denotes the distance between user groups $g$ and $f$, $d_0$ the cutoff distance,
$\alpha$ the path-loss exponent, $w$ the wall penetration loss, and $n_w(g,f)$ denotes the number of walls between
user groups $g$ and $f$. We have
\begin{equation}
n_w(g,f) = \left\{\begin{array}{ll}
0 & g=f \\
1 & g \in \Mc, f \in \Sc \ {\rm OR\ } g \in \Sc, f \in \Mc\\
2 & g , f \in \Sc, g\neq f
\end{array}\right..
\end{equation}

\subsection{Received signals}

In this work, we focus on the downlink of both the macrocell as well as the small cell tiers.
As discussed in Section~\ref{subsec:macro}, macrocells are typically deployed at some elevation, which means they have a fairly narrow
angular spread to each user group. The resulting channel vectors are therefore correlated, with covariances given (in our model) by (\ref{one-ring}).
Such directional information can be exploited in order to simplify the multiuser MIMO beamforming. In particular, in this work we consider to use
JSDM,
a two stage beamforming scheme proposed in \cite{adhikary2012joint} (see also \cite{adhikary2013joint,adhikary2013joint2}) in order
to achieve massive-MIMO like gains with reduced feedback overhead requirements for channel state information and complexity in terms of the number of
baseband-to-RF chains.
The idea is to partition the user space into groups of users with approximately similar covariances,
and split the downlink beamforming into two stages: a first stage consisting of a pre-beamformer that depends
only on the second order statistics, i.e., the covariances of the user channels, and a second stage comprising a standard multiuser MIMO precoder for
spatial multiplexing on the effective channel including the pre-beamforming.
The instantaneous channel state information for such scheme  is easier to acquire because of the considerable dimensionality reduction of the effective channel
produced by the pre-beamforming stage. In addition, JSDM lends itself to a hybrid beamforming implementation, where pre-beamforming
may be implemented in the analog RF domain, while the multiuser MIMO precoding stage is implemented by baseband processing.
This approach allows the use of a very large number of antennas with a limited number of baseband-to-RF chains, that
depends on the number of independent data streams sent simultaneously to the users, and not on the number of macrocell antennas $M$, which can be
made very large.

Recall that the macrocell serves $G$ user groups using JSDM. The received signal vector $\yv_g$ of the users located in group $g$ can be expressed as
\begin{eqnarray} \label{eqn:mc}
\yv_g
% & = & \Hm_{g,0}^\herm \Vm \dv + \sum_{f \in \Sc_A} \Hm_{g,f}^\herm \Qm_f \sv_f + \zv_g \nonumber\\
&=& \Hm_{g,0}^\herm \Bm_g \Pm_g \dv_g
+ \sum_{g' \neq g} \Hm_{g,0}^\herm \Bm_{g'} \Pm_{g'} \dv_{g'}
+ \sum_{f \in \Sc_A} \Hm_{g,f}^\herm \Qm_{f,f} \sv_f + \zv_g,
\end{eqnarray}
%where $\Vm = \left[\Bm_1 \Pm_1 \ldots \Bm_G \Pm_G\right]$
where $\dv_g$ is the $S_g \times 1$ vector of transmitted data symbols to the group $g$ users, $\Pm_g \in \CC^{b_g \times S_g}$ and $\Bm_g \in \CC^{M \times b_g}$ are the precoding and pre-beamforming matrices\footnote{The prebeamforming dimension $b_g$ determines the amount of channel state information to be fed back to the transmitter and should be carefully optimized (see \cite{adhikary2012joint} for details).} for group $g$ of the JSDM scheme, and  $\Hm_{g,0} = \left[ \hv_{g_1,0} \ldots \hv_{g_{S_g},0} \right]$ is the channel matrix between the macrocell antenna array and the served users in group $g$. Notice that the structure of the JSDM precoder, split into the product $\Bm_g \Pm_g$  is not redundant, since
we impose (by JSDM system design constraint) that $\Bm_g$ depends only on the channel second-order statistics information
$\{\Rm_g: g = 1, \ldots, G\}$, while $\Pm_g$ is allowed to depend on the instantaneous realization of the projected channels
$\{\Hm_{g,0}^\herm \Bm_g : g = 1, \ldots, G\}$.
The matrix $\Hm_{g,f} = \left[ \hv_{g_1,f} \ldots \hv_{g_{S_g},f} \right]$ contains the channels between users in group $g$ and small cell $f$,
$\Qm_{f,f}$ is the precoding vector used by small cell $f$, $\sv_f$ is the vector of data symbols transmitted by small cell $f$, $\Sc_A$ is the set of active small
cells resulting from the various inter-tier interference coordination strategies discussed in Section \ref{sec:inter-tier-ic}, and $\zv_g$ denotes the additive white Gaussian noise, with i.i.d. components $\sim \Cc\Nc(0,1)$.

We assume that the macrocell uses JSDM with {\em per group processing (PGP)} (see \cite{adhikary2012joint}),
since this has the advantage of significantly reducing the channel state information feedback requirement and also ensures
that the second stage precoding matrix $\Pm_g$ can be independently designed across all groups $g$.
PGP results in an additional inter-group interference term, given by the sum over $g' \neq g$ in (\ref{eqn:mc}),
which can be eliminated by block diagonalization \cite{adhikary2012joint,spencer2004zero} or by serving groups of users with disjoint
angular support using DFT prebeamforming, in the limit of very large $M$ (see details in \cite{adhikary2012joint}).
With PGP, the precoding matrix $\Pm_g$ depends only on the instantaneous effective channel $\Bm_g^\herm \Hm_{g,0}$.
We consider zero forcing beamforming, such that $\Pm_g$ is given by the Moore-Penrose pseudoinverse of $\Bm_g^\herm \Hm_{g,0}$, up to a power normalization scalar factor. Specifically, we have
\begin{equation}
\Pm_g = \zeta_g \Bm_g^\herm \Hm_{g,0} \left( \Hm_{g,0}^\herm \Bm_g  \Bm_g^\herm \Hm_{g,0} \right)^{-1},
\end{equation}
where $\zeta_g$ is the normalization factor. Notice that (\ref{eqn:mc}) also includes a term that captures the interference that all the
active small cells (in the set $\Sc_A$) cause to the group $g$ users.

Since the small cells  are equipped with $L$ antennas each, they can serve up to $L$ users.
We assume that the number of active users, denoted by $\bar{S}_f$,  is the same in all the small cells,
and is equal to a certain fraction of number of antennas $L$, i.e., we let $\bar{S}_f = \bar{S} = \beta L \ \forall \ f \in \Sc$,
where $\beta$ is a design parameter that depends on the precoding scheme. Small cells make use of  zero forcing precoding to serve their users,
so that the received signal of users in group $f$ served by its corresponding small cell is given by
\begin{eqnarray} \label{eqn:sc}
\bar{\yv}_f &=& \Hm_{f,f}^\herm \Qm_{f,f} \sv_f
+ \sum_{f' \neq f, f' \in \Sc_A} \Hm_{f,f'}^\herm \Qm_{f',f'} \sv_{f'}
+ \sum_{g=1}^G \Hm_{f,0}^\herm \Bm_g \Pm_g \dv_g + \bar{\zv}_f
\end{eqnarray}
where $\Hm_{f,f'} = \left[ \hv_{f_1,f'} \ldots \hv_{f_{\bar{S}},f'} \right]$ is the channel matrix for the users in group $f$ and the small cell array of group $f'$,
$\Qm_{f,f}$ is the zero forcing precoding matrix of small cell $f$ (given as the column-normalized Moore Penrose pseudo inverse of $\Hm_{f,f}$)
and $\bar{\zv}_f$ is the additive white Gaussian noise. The users in group $f$ suffer interference from all the other active small
cells (sum over $f' \neq f$ in (\ref{eqn:sc})), along with interference from the macrocell (sum over $g$ in (\ref{eqn:sc})).
We also assume that the small cells transmit at peak power $P_1$, and all users' data stream are allocated equal power $P_1/\bar{S}$.

%%%%%%%%%%%%%%%%%%%%%%%%%%%%%%%%%%%%%%%%%%%%%%%%%
\subsection{Expressions for Received SINR} \label{sec:da-sinr}

From (\ref{eqn:mc}), the received SINR at a macrocell user $k$ in group $g$ is given by
\begin{equation}
{\rm SINR}_{g_k}^{\rm mc} = \frac{|\hv_{g_k,0}^\herm \Bm_g \pvx_{g_k}|^2 \frac{P_0}{S}}{1 + \sum_{g' \neq g} || \hv_{g_k,0}^\herm \Bm_{g'} \Pm_{g'} ||^2 \frac{P_0}{S} + \sum_{f \in \Sc_A} || \hv_{g_k,f}^\herm \Qm_{f,f} ||^2 \frac{P_1}{\bar{S}}}.
\end{equation}
Using results from random matrix theory, we approximate the ${\rm SINR}_{g_k}$ using the techniques of ``deterministic equivalents'' (\cite{wagner2012large}, \cite{adhikary2012joint}) by a quantity ${\rm SINR}_g^{\rm mc,DE}$, which is common to all users being served in group $g$, where the approximation holds almost surely when the number of antennas $M \longrightarrow \infty$. This is facilitated by the assumption that all the users in a given group have the same channel covariance matrix. For completeness, we provide the expression for ${\rm SINR}_g^{\rm mc, DE}$.
\begin{equation}
{\rm SINR}_g^{\rm mc, DE} = \frac{D_{g,0}^{\rm mc} \frac{P_0}{S}}{1 + \sum_{g' \neq g} I_{g,g'}^{\rm mc} S_{g'} \frac{P_0}{S} + \sum_{f \in \Sc_A} J^{\rm sc}_{g,f} P_1}
\end{equation}
where $D_{g,0}^{\rm mc} = b_g m_g$, and $m_g$ is given by the solution of the following fixed point equation
\begin{eqnarray}
m_g &=& \frac{1}{b_g} {\rm trace} \left( \Bm_g^\herm \Rm_g \Bm_g \Tm_g^{-1} \right) \nonumber\\
\Tm_g &=& \Id_{b_g} + \frac{S_g}{b_g} \frac{\Bm_g^\herm \Rm_g \Bm_g}{m_g}
\end{eqnarray}
and
\begin{eqnarray}
I^{\rm mc}_{g',g} &=& \frac{n_{g',g}}{m_g} \label{eqn:interf-mc} \\
n_{g',g} &=& \frac{\frac{1}{b_g} {\rm trace} \left( \Bm_{g}^\herm \Rm_{g} \Bm_{g} \Tm_{g}^{-1} \Bm_{g}^\herm \Rm_{g'} \Bm_{g} \Tm_{g}^{-1} \right)}{1 - F_g} \label{eqn:defn}\\
F_g &=& \frac{1}{b_{g}} \frac{\frac{S_{g}}{b_{g}} {\rm trace} \left( \Bm_{g}^\herm \Rm_{g} \Bm_{g} \Tm_{g}^{-1} \Bm_{g}^\herm \Rm_{g} \Bm_{g} \Tm_{g}^{-1} \right)}{m_g^2} \nonumber \\
J^{\rm sc}_{g,f} &=& a_{g,f} \label{eqn:interf-sc-mc}
\end{eqnarray}

Similarly, from (\ref{eqn:sc}), the expression for the received SINR at a user $k$ in group $f$ being served by the small cell in user group $f$ is given by
\begin{equation}
{\rm SINR}_{f_k,f}^{\rm sc} = \frac{|\hv_{f_k,f}^\herm \qv_{f_k,f} |^2 \frac{P_1}{\bar{S}}}{1 + \sum_{g = 1}^G || \hv_{f_k,0}^\herm \Bm_{g} \Pm_{g} ||^2 \frac{P_0}{S} + \sum_{f' \in \Sc_A, f' \neq f} || \hv_{f_k,f'}^\herm \Qm_{f',f'} ||^2 \frac{P_1}{\bar{S}}}
\end{equation}
The deterministic equivalent approximation for the quantity ${\rm SINR}_{f_k,f}^{\rm sc}$ is given by ${\rm SINR}_{f,f}^{\rm sc, DE}$ such that
\begin{equation}
{\rm SINR}_{f,f}^{\rm sc, DE} = \frac{D_{f,f}^{\rm sc} \frac{P_1}{\bar{S}}}{1 + \sum_{g=1}^G J^{\rm mc}_{f,g} S_g \frac{P_0}{S} + \sum_{f' \in \Sc_A,f' \neq f} I^{\rm sc}_{f,f'} P_1}
\end{equation}
where $D_{f,f}^{\rm sc} = a_{f,f} (L - \bar{S} + 1)$, $I^{\rm sc}_{f,f'} = a_{f,f'}$ and
$$J^{\rm mc}_{f,g} = \frac{n_{f,g}}{m_g},$$ with $n_{f,g}$ defined in (\ref{eqn:defn}).

%%%%%%%%%%%%%%%%%%%%%%%%%%%%%%%%%%%%%%%%%%%%%%%%%%%%%%%%%
%%%%%%%%%%%%%%%%%%%%%%%%%%%%%%%%%%%%%%%%%%%%%%%%%%%%%%%%%
\section{Inter-Tier Coordination Strategies} \label{sec:inter-tier-ic}

In this section, we focus on various inter-tier coordination strategies in order to reduce interference experienced by both the macro and small cell user groups.
Recall that the user groups which are not covered by a small cell can either be served by the macrocell
using JSDM or ``offloaded'' to a nearby small cell.
Similarly, the ones that do have a co-located small cell, have two options: (i) the small cell is turned off
depending on the amount of interference it causes to the nearby macrocell user groups; or
(ii) the macrocell does not form a beam in that direction allowing the small cell to be active and serve users in its group with good SINR.
Since we assume random locations of the user groups, this might lead to several conflicts, e.g., there may be
two user groups with similar angle of arrival with respect to the macrocell, one covered by a small cell and the other with no dedicated small cell.
In this case, the small cell may shut down depending on the amount of interference that it receives from the macrocell
while the user group without dedicated small cell  is being served by the macrocell.
On the other hand, the user groups may be located too close to each other. In this case, it may be advantageous to just let the small cell of one group serve both groups,
while the macrocell does not form beams in that direction. %An example is shown in Fig. \ref{fig:inter-tier}.
Four explicit schemes that achieve such inter-tier interference coordination are outlined in the rest of this section.
%Specifically, we consider four different strategies and compare their performance in Section \ref{sec:numerical-results}. Each strategy has its own relative merits
%depending on the overall objective, and the strategies compare differently depending on the type of small cell deployment (see Definition \ref{def:smallcellscenarios}).

\subsection{No Coordination} \label{algo:usr-grp-sel}

As the name suggests, in this scheme, we do not assume any explicit inter-tier coordination between the macrocell
and the small cells.
In each transmission resource block, the macrocell selects and serves a subset of size $G$ of user groups from the set $\Mc$, according to some fair group selection algorithm ensuring
that every user group in the set $\Mc$ is given equal air time. Furthermore, the group selection algorithm checks that the set of groups served simultaneously
has channel covariance eigenspaces {\em approximately} mutually  orthogonal.\footnote{More advanced schemes that guarantee more general notions of fairness can be formulated as a weighted sum rate maximization problem using tools from stochastic
network optimization (see \cite{shirani2010mimo} and references therein), and by defining a suitable concave network utility function of the user long-term averaged rates.
Such schemes give a performance guarantee that depends on a parameter $\delta$, where the the convergence time grows as
$O(\delta)$ in order for the scheme to perform within a gap $O\left(\frac{1}{\delta}\right)$ from the optimal value of the network utility function.}
Our user group selection algorithm does not take into account the small scale fading statistics and makes decisions based only on the second order statistics
of the user groups, thus eliminating the need for explicit channel state feedback from all the users at every scheduling slot.
The feedback is only required to design the PGP precoder $\Pm_g$ for every group $g$, which
depends on the instantaneous effective channel $\Bm_g^\herm \Hm_{g,0}$, as already remarked before.
We now outline the user group selection algorithm.

{\em Algorithm for user group selection:} The algorithm works by maintaining a priority vector for every user group in set
 $\Mc$, i.e., the user groups that do not contain a small cell, and updating the priority vector so as to maintain a high priority for user groups that have not been served in the recent time slots. We first initialize the priority of all user groups to 1 and,
 at the end of the scheduling slot, we increment the priorities of the non selected user groups by 1, while keeping the priorities of the selected user groups unchanged.
 This guarantees an equal opportunity to all the user groups to be served by the macro BS, so that no user group is left starving.
 Also, using the results from Szego's theory on large dimensional Toeplitz matrices (see \cite{adhikary2012joint}), we have that the channel covariance eigenspaces of two user groups are approximately
 orthogonal when their channel angular support intervals are disjoint.
 Therefore, at every scheduling slot, the macrocell greedily selects a user group which has the highest priority and at the same time, causes the least interference
 to the already selected user groups.
 \begin{itemize}
 \item {\bf Step 1:} Given the user group priority vector $\cv$, initialize $\Gc = g^*$, $\Mc_{\rm res}^{(0)} = \Mc \setminus g^*$, $S^{(0)} = \sum_{g \in \Gc} S_g = S_{g^*}$, where $g^*$ is chosen randomly from the set of user groups in $\Mc$ with highest priority. $\Mc_{\rm res}$ denotes the set of user groups that are compatible (in terms of their channel angular support)  with the already selected user groups. We define a set of intervals $\Ic_g$ for every user group $g$, which is a function of the user group's angles of arrival $\theta_g$ and angular spread $\Delta_g$ and is given by $$\Ic_g = \left[ -\frac{1}{2} \sin(\theta_g + \Delta_g), -\frac{1}{2} \sin(\theta_g - \Delta_g) \right]$$
 \item {\bf Step 2:} At iteration $n$, we define a set $\Mc_{\rm res}^{(n)} = \emptyset$, and add only those user groups to this set which have non overlapping intervals with the already selected user groups in $\Gc$. Thus, for every user group $g' \in \Mc_{\rm res}^{(n-1)}$,
 $$\Mc_{\rm res}^{(n)} = \Mc_{\rm res}^{(n)} \bigcup g' \ \ {\rm if} \ \ \Ic_{g'} \bigcap \Ic_g = \emptyset \ \ \forall g \in \Gc$$
 \item {\bf Step 3:} We now select a user group $g$ that has the highest priority and minimizes the maximum interference to the already selected user groups in $\Gc$. Since $|| \hv_{g_k,0}^\herm \Bm_{g'} \Pm_{g'} \Pm_{g'}^\herm \Bm_{g'}^\herm \hv_{g_k,0} ||^2 = S_{g'} I^{\rm mc}_{g,g'}$ is the inter-group interference between a user $k$ in group $g$ and the users in group $g'$ (note that $I^{\rm mc}_{g,g'}$ defined in (\ref{eqn:interf-mc}) is independent of the fading realizations $\hv_{g_k,0}$, using results from asymptotic random matrix theory),
 the total inter-group interference to a user group $g$ from all the precoded data streams sent to users in group $g'$
 is given by $\frac{P_0 S_{g'}}{S^{(n-1)} + S_{g'}} I^{\rm mc}_{g,g'}$. For every $g' \in \Mc_{\rm res}^{(n)}$, we compute
     $$I_{g'}^{\max} = \max_{g \in \Gc} \frac{P_0 S_{g'}}{S^{(n-1)} + S_{g'}} I^{\rm mc}_{g,g'} $$
 \item {\bf Step 4:} Assign $g^*$ to $\Gc$ which has the maximum priority and minimizes the maximum interference to the already selected user groups. Define $c_{\max}$ as the maximum element of the vector $\cv$ and a set $\Cc = \emptyset$. For every $g' \in \Mc_{\rm res}^{(n)}$,
     $$\Cc = \Cc \bigcup g' \ \  {\rm if} \ \ c_{g'} = c_{\max}$$
 \item {\bf Step 5:} Find the user group $g^*$ such that $$g^* = {{\rm arg} \min}_{g' \in \Cc} I_{g'}^{\max}$$ and update
 $$\Gc = \Gc \bigcup g^*, \ \ S^{(n)} = \sum_{g \in \Gc} S_g, \ \ \Mc_{\rm res}^{(n)} = \Mc_{\rm res}^{(n)} \setminus g^*$$
 \item {\bf Step 6:} If $\Mc_{\rm res}^{(n)} = \emptyset$ or $|\Gc| = G$, stop and output $\Gc$ as the result, else increment $n$ by 1 and go to Step 2.
 \end{itemize}

After the selection of user groups, the macrocell selects (uniformly at random) $S_g$ users from each selected group $g$, and serves them using JSDM with PGP and zero-forcing precoding in each group to eliminate the intra-group interference as explained before. From the results in \cite{adhikary2012joint}, for large number of antennas $M$ and covariance group rank $r_g$, it is known that the optimal value of $S_g$ is given by $\beta r_g$ for some design parameter $\beta < 1$
that can be optimized depending on the scattering geometry.  In order to keep the problem tractable and obtain meaningful results, in this paper we use the same value of $\beta$ for all groups $g$.

\begin{remark}
In implementing the user selection algorithm, we set the pre-beamforming matrix $\Bm_g$ of every user group being served by the macrocell as $\Bm_g = \Um_g^*$,
where $\Um_g^*$ contains the eigenvectors corresponding to the dominant eigenvalues of the corresponding channel covariance matrix $\Rm_g$.
This greatly simplifies the algorithm as opposed to performing block diagonalization (or approximate block diagonalization), which requires recomputing the prebeamformers every time a new user group is selected in the scheduled pool of the
macrocell. Such a simplification does not come to an overly pessimistic performance price when the number of antennas $M$ is very large, since in this case
the channel covariances of two user groups with disjoint angular support are approximately orthogonal to each other and therefore, by virtue of  Step 2 of the algorithm,
(approximate) block diagonalization is implicitly achieved.
\end{remark}

In the {\em no coordination} scheme,  the small cells just transmit to their own users using zero forcing beamforming at their own peak total power. Recall that small cells are equipped with $L$ antennas, and they serve $\bar{S} = \beta L$ users in every transmission resource block, selected at random in order to give equal air time to all their users. In this case,  all the small cells are active on all transmission resource blocks. However, it is interesting to see that when the number of user groups $G$ served by the macro BS is not too large, thanks to the inherent directionality in pre-beamforming achieved by JSDM, the macrocell implicitly mitigates the interference at the small cells that are not aligned in the direction of the pre-beamforming vectors. Thus, compared with a naive uncoordinated scheme that serves users isotropically instead of co-located user groups, our uncoordinated scheme is able to achieve some non-trivial interference suppression benefits. We illustrate this point in the following toy example.

\begin{figure}[!h]
  \centering
\begin{tikzpicture}[scale=.5]
 \node[anchor=south west,inner sep=0] at (-10,-4.5) {\includegraphics[width=.49\textwidth]{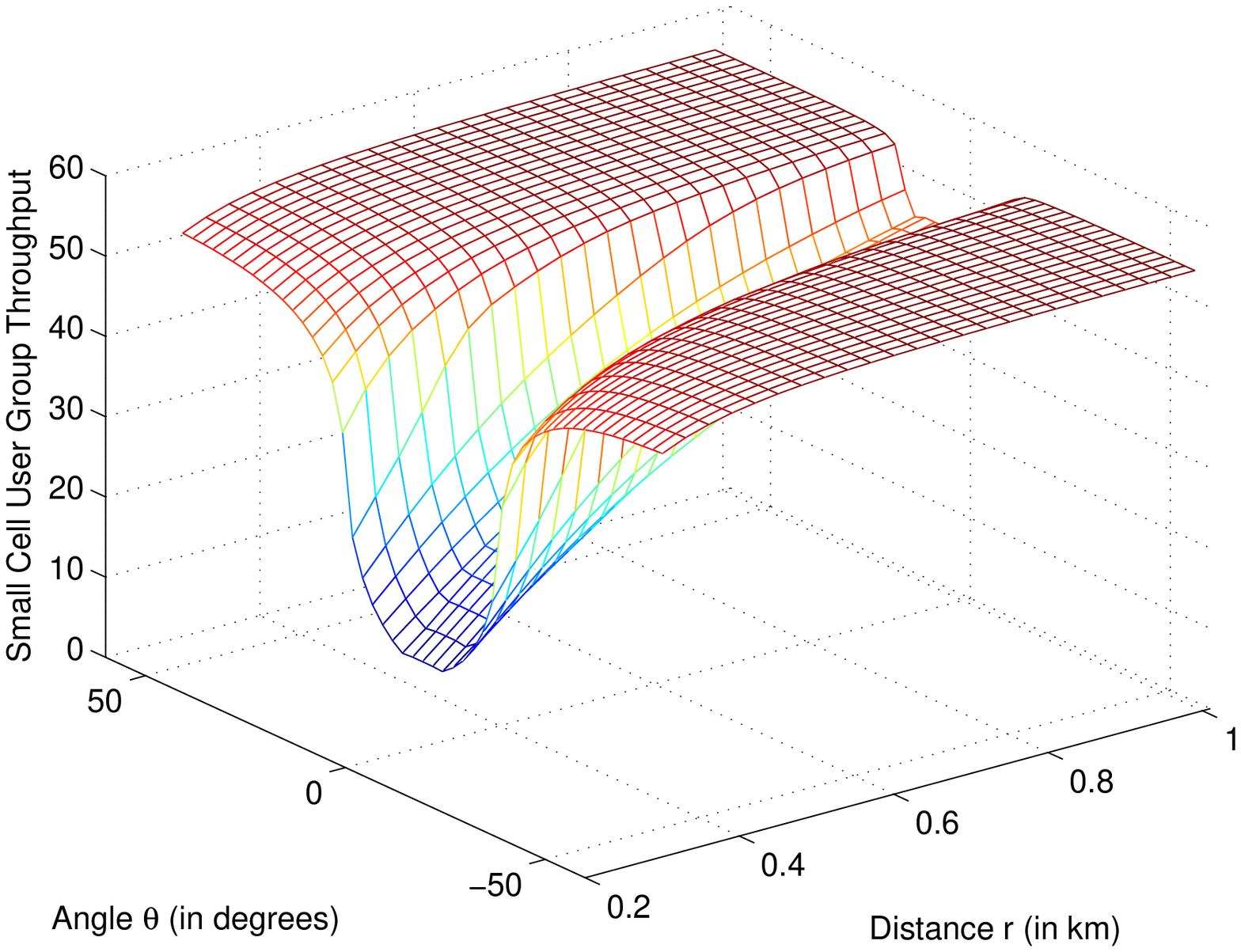}};
\draw (-1,0) -- (5,0);
\draw (0,-2) -- (0,2);
\draw (2,0) [blue, ultra thick] circle [radius = 0.4];
%\node [text width = 2.5cm] at (3,-1.5) {\scriptsize Macrocell User Group};
\draw[dashed, thick] (0,0) -- (3,0.6124);
\draw[dashed, thick] (0,0) -- (3,-0.6124);
\draw (4,1.5) [fill=red,red] circle [radius = 0.4];
%\node [text width = 2.5cm] at (4,2) {\scriptsize Small Cell User Group};
\draw [->, thick] (0,0) -- (4,1.5);
\node at (2,1) {\scriptsize {$r$}};
\draw [->,thick] (3.5,0) arc [radius=3.5, start angle = 0,end angle = 20.5];
\node at (3.7,0.6) {\scriptsize{$\theta$}};
\end{tikzpicture}
  \includegraphics[width=0.49 \textwidth]{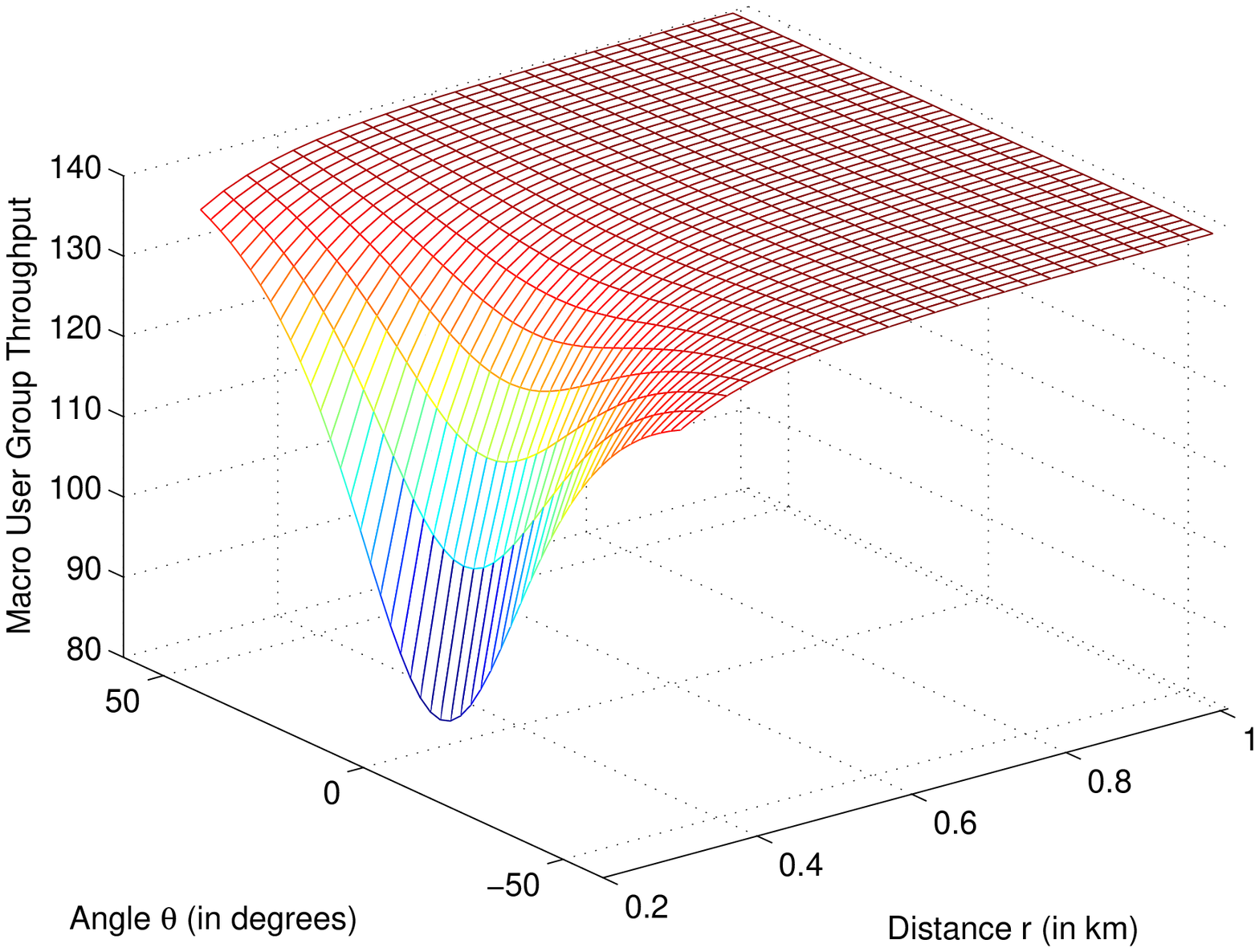}
  \caption{The toy example to demonstrate spatial blanking in Example~\ref{example1}. {(\em first)} The throughput of the small cell user group. {\em (second)} The throughut of the macrocell user group. The inset in the first figure shows the layout used in this example. The hollow and filled circles denote macro and small cell user groups, respectively.}
  \label{fig:example}
\end{figure}

\begin{example}[Interference suppression in {\em no coordination} strategy]  \label{example1}
Consider a toy example with two active user groups: one served by the macrocell and the other by a small cell. The macro user group is assumed to be located at a distance of 0.2 km from the macrocell. The location of the small cell user group is parameterized by $(r,\theta)$ as shown in the inset of the first figure of Fig.~\ref{fig:example}. Note that $\theta=0$ means that both the user groups are aligned when seen from the macrocell, which results in a high cross-tier interference at the small cell user group. As $|\theta|$ is increased, this interference almost vanishes leading to high small cell throughput (first figure of Fig.~\ref{fig:example}). This is precisely what we mean by ``spatial blanking'' in this paper. Also note that the distance between the two user groups dictates the cross-tier interference seen at the macrocell user group (second figure of Fig.~\ref{fig:example}). When this distance is reduced, the macro user rate drops significantly, which necessitates explicit interference coordination strategies to complement spatial blanking, which we do in the rest of this section. The simulation parameters used for Fig.~\ref{fig:example} are the same as the ones used for the numerical results in Section~\ref{sec:numerical-results}. They are tabulated in Table~\ref{table:sim-param}.
\end{example}

\subsection{Coordination Scheme 1: ON/OFF} \label{subsec:on-off}

%In the uncoordinated scheme, no constraint is imposed over the interference that the small cells cause to the macrocell users.
%In fact, since all the small cells are active on every resource block, these would create significant interference to a user group being served
%by the macrocell and would therefore degrade its rate. Also, due to the random placement, the small cells  close to the macrocell and aligned with the direction of the pre-beamforming vectors of the macrocell
%would suffer from large interference, resulting in performance degradation.
In the coordination scheme of this section, the macrocell first chooses a set of user groups to be served using the same user group selection algorithm of Section \ref{algo:usr-grp-sel}. The small cells then use this information and the knowledge of the cross-tier interference and the useful signal strengths to implement a simple ON/OFF strategy for the
given scheduling slot. Explicitly, a small cell decides to shut down its transmission based on the amount of cross-tier interference it receives or causes to the scheduled macrocell user groups.
In order to make this decision, the small cell compares the interference that its users receive  from the macrocell with the strength of their useful signal,
as well as the interference that it causes to the macrocell user groups with the strength of the useful signal at these groups.
We formally present the strategy next.

For a user $k$ in group $f$ that is being served by a small cell, the amount of interference caused by the macrocell is given by $J^{\rm mc}_{f,0} = \sum_{g=1}^G || \hv_{f_k,0}^\herm \Bm_g \Pm_g ||^2 = \sum_{g=1}^G S_g J^{\rm mc}_{f,g}$.
Similarly, the useful signal strength, as a result of zero forcing beamforming is given by $D^{\rm sc}_{f,f} = || \hv_{f_k,f}^\herm \qv_{f_k,f} ||^2$.
The useful signal strength of a user $m$ in group $g$ served by the macrocell  is given by $D^{\rm mc}_{g,0} = || \hv_{g_m,0}^\herm \Bm_g \pvx_{g_m} ||^2$ and the
interference caused by small cell $f$ to this particular user is given by $|| \hv_{g_k,f}^\herm \Qm_f ||^2 = \bar{S} J^{\rm sc}_{g,f} $.
Expressions for $J^{\rm mc}_{f,0}$, $D^{\rm sc}_{f,f}$, $J^{\rm sc}_{g,f}$ and $D^{\rm mc}_{g,0}$ are given in Section \ref{sec:da-sinr}.
Using these quantities, we can formulate the proposed ON/OFF strategy for the small cells.

A small cell serving a user group $f$ decides to transmit or shut down according to the following simple criterion
\begin{equation}
{\rm Small\ Cell\ }f = \left\{\begin{array}{cc}
{\rm ON} & \frac{P_0}{S} J^{\rm mc}_{f,0} \leq \epsilon_1 D^{\rm sc}_{f,f}  \frac{P_1}{\bar{S}} \ {\rm and} \ J^{\rm sc}_{g,f} P_1 \leq \epsilon_2 D^{\rm mc}_{g,0} \frac{P_0}{S} \ \forall \ g \in \Gc \\
{\rm OFF} & {\rm otherwise}
\end{array}\right.
\end{equation}
where $\epsilon_1$ and $\epsilon_2$ are design parameters that can be set to achieve a desired tradeoff between
the macrocell and small cell throughputs.

\subsection{Coordination Scheme 2: OFFLOAD} \label{subsec:offload}

In the ON/OFF coordination strategy, the cross-tier interference is mitigated by the small cells deciding on their transmissions.
This strategy may be disadvantageous to a user group being served by a small cell located in the vicinity of another user group not covered by a small cell.
This is because whenever the macrocell schedules to serve such an uncovered user group, the small cell in the neighboring covered group has to shut down.
A simple way to alleviate this problem consists of associating
some of these ``bottleneck'' user groups devoid of small cell to the nearest small cell,
so that the small cell  can now serve both user groups using TDMA.
We call this the ``OFFLOAD'' strategy, because some of the macrocell user groups are being offloaded to neighboring small cells in order to be able to be served more efficiently and cause less harm
in terms of their imposed interference constraints.  It is worthwhile to notice that this strategy is intended to increase the throughput of macrocell user groups which are located at the cell edge and have a small cell close to them.
This is because the macrocell user groups at the cell edge receive a low useful signal strength from the macrocell and, in addition, are likely to block the transmission of the neighboring small cells.
However, the offloading approach results in a decrease of the rates observed in the small cells,
because of the fact that the small cells may serve two or more user groups in TDMA, thereby reducing the average throughput
of its own users by a factor equal to the number of associated user groups.
Formally, a small cell $f$ decides upon absorbing a user group $g$ which normally would have been served by the macrocell  according to the following condition:
\begin{equation}
\begin{array}{cc}
D^{\rm sc}_{g,f} \frac{P_1}{\bar{S}} > \gamma D^{\rm mc}_{g,0} \frac{P_0}{S_g} & {\rm OFFLOAD}\\
D^{\rm sc}_{g,f} \frac{P_1}{\bar{S}} \leq \gamma D^{\rm mc}_{g,0}  \frac{P_0}{S_g} & {\rm NO\ OFFLOAD},
\end{array}
\end{equation}
where $D^{\rm sc}_{g,f} = || \hv_{g_k,f}^\herm \qv_{g_k,f} ||^2$ and $D^{\rm mc}_{g,0} = || \hv_{g_k,0}^\herm \Bm_g \pvx_{g_k} ||^2$, where $D^{\rm sc}_{g,f}$
is the direct link gain between a small cell $f$ and a user $k$ in group $g$ and $D^{\rm mc}_{g,0}$ is the direct link gain of the same user from the macrocell, assuming that the macro BS serves the user group $g$ in isolation.
The parameter $\gamma$ moderates the fraction of macrocell user groups being offloaded, e.g., a small value of $\gamma$ indicates that more macrocell user groups will be offloaded to the small cells.
We denote the set of offloaded user groups as $\Mc_{\Sc}$.

It is worth mentioning that the offloading strategy also reduces the burden on the small cells compared to the ON/OFF strategy, where the small cells were required to make decisions on every scheduling slot by knowing the cross-tier interference.
In fact, in this case, the small cells can decide in advance the user groups to serve and relay this information to the macrocell, so that the latter does not include these user groups in its scheduling selection.
Such information must be conveyed on a time scale of a few seconds (assuming low mobility, we can safely assume that the distribution of the user groups remains static over long periods of time when compared to the time slot for making scheduling decisions) without any significant protocol overhead.

\subsection{Coordination Scheme 3: TIN} \label{subsec:tin}

In this section, we propose an inter-tier interference coordination scheme based on a recent result in \cite{geng2013optimality},
that gives conditions on the (approximated) optimality of treating interference as noise (TIN)
in a Gaussian interference channel, i.e., a network consisting of several mutually interfering links.
The conditions depend on the channel gains of both the direct and the interfering links, normalized by a common transmit power factor.
In our setup consisting of a single macrocell and several small cells  with multiple antennas, we can view the network as a set of links, where each link corresponds to a particular user being
served and the direct and cross link gains are a function of the transmit power allocated to the user data streams as well as of the beamforming vectors.
The number of links is equal to the total number of user data streams transmitted by both the macrocell and the small cells. Also, owing to the deterministic equivalent approximations
to the SINRs given in Section \ref{sec:da-sinr}, the gains of both the direct and cross  links of the users belonging to a particular group are identical (since they do not depend on the instantaneous fading realizations, but only on the channel covariance structure).
We first formally state the TIN optimality condition and then present our algorithm to find a set of user groups to be served on a given scheduling slot that satisfy
the TIN optimality condition.

%%%%%%%%%%%%%%%%%%%%%%%%%%%%%%%%%%%%%%%%%%%%%%%%

Given a set of links $\{1,2,\ldots,N\}$ with transmit power $(P_1,P_2,\ldots,P_N)$ and direct link gains $\xi_{ii}$ and cross link gains $\xi_{ij}$, the condition for these set of links
to be TIN optimal is given as
\begin{equation} \label{eqn:tin-opt-condn}
\xi_{ii} P_i \geq \left[ \max_{j \neq i} \xi_{ij} P_i \right] \times \left[ \max_{j \neq i} \xi_{ji} P_j \right] \ \forall \ i \in \{1,2,\ldots,N\}
\end{equation}
For a given network of interfering links, finding a set of maximal size that satisfies the TIN optimality condition has a worst case complexity that is polynomial in the number of links. However, finding a set that is maximal and optimizes a desired objective is in general hard \cite{naderializadeh2013itlinq}. We therefore propose the ``TIN Selection Algorithm'', which is a greedy heuristic for choosing a set of user groups so that the set of scheduled links in the network is of maximal size and satisfies the TIN optimality condition. We present the main idea of the algorithm in the following paragraph, and defer the details to Appendix \ref{appendix-tin-algo}.

We start by forming a set of user groups that have the highest priority using the same procedure as outlined in Section~\ref{algo:usr-grp-sel} for the uncoordinated strategy. From this set, we then choose a user group that has the highest direct link gain. Then, at every iteration, we check over all the remaining user groups that have not been selected to see whether the addition of a user group to the already selected user groups would violate the TIN optimality condition or not. Note that the addition of a macrocell user group changes the power allocated to every active link being served by the macrocell, since we assume that the macrocell power is fixed and all the served macrocell user streams are allocated equal power. Only those user groups whose addition would not violate the TIN optimality conditions are kept in a residual set. From this set, a user group with the highest priority is chosen according to a heuristic, which is described in Steps 5 and 6 of the algorithm in Appendix \ref{appendix-tin-algo}. This process is continued until there are no more user groups remaining in the residual set. After having a set of selected user groups, the priority values for the non-selected user groups are incremented by 1, while those of the selected ones are kept unchanged. The TIN selection algorithm, in Steps 2 and 3 always adds only those user groups that satisfy the TIN optimality conditions with respect to the already selected user groups. Therefore, it ensures that the final output of the selected user groups satisfy the TIN optimality condition. Also, the algorithm terminates when no more user groups can be added without violating the TIN optimality conditions, which indicates that the resulting set is maximal, although it may not be the ``optimal'' set.

%
%
%The algorithm ensures that the resulting selected set of user groups satisfy the TIN optimality condition.

%\chr{Again we need to provide some intuition for each step before providing the expressions.}

%\begin{figure}[!h]
%  \centering
%  % Requires \usepackage{graphicx}
%  \includegraphics[width=10cm]{inter-tier}\\
%  \caption{An example showing inter tier coordination.}\label{fig:inter-tier}
%\end{figure}

%\section{Performance Analysis}
%\chb{If it improves the presentation, you could first assume only multi-antenna small cells, i.e., a two-tier scenario and then write a separate subsection extending the 2-tier analysis to 3-tiers.}
%{Once we have cleared everything about the model and the system assumptions, we can proceed with the analysis.}

%{\RED This should not be a problem since the results are mostly known and we just need to apply them once we have the model in hand.}

\section{Numerical Results} \label{sec:numerical-results}

In this section, we present numerical results and discuss the relative merits of inter-tier coordination strategies developed in the paper. The default simulation parameters are listed in Table \ref{table:sim-param}. We present results for $N_{\rm f} = 20$ and $N_{\rm f} = 50$, corresponding to a low and high density of small cells, respectively. For each case, we further consider three ways the small cells can be deployed: (i) uniform, (ii) cell-interior, and (iii) cell edge, which were formally defined in Definition~\ref{def:smallcellscenarios}. Unless stated otherwise, in all the figures, the results corresponding to the uniform, cell-interior and cell-edge deployments are presented using dash-dotted, dashed and solid lines respectively.  We consider an exclusion ball of radius $R_{\rm excl}$ around the macrocell which does not contain any user groups. The peak power of the small cells is set to $P_1 = \frac{P_0}{100}$, corresponding to a value that is 20 dB less than the peak power used by the macrocell. The macrocell power $P_0$ is calculated from the cell edge SNR value given in Table \ref{table:sim-param}.

\renewcommand{\arraystretch}{0.6}
\begin{table}[t]
 \caption{List of simulation parameters}
\centering
 \begin{tabular}{|c|c|}
 \hline Parameter & Value \\
 \hline No. of user groups $N_{\rm u}$ & 500\\
 \hline  Cell radius $R_{\rm mc}$ & 1 km \\
 \hline  Cut off distance $d_0$ & 50 m \\
 \hline  Exclusion radius $R_{\rm excl}$ & 100 m \\
  \hline Path loss exponent $\alpha$ & 3.5 \\
 \hline Cell edge SNR & 10 dB \\
 \hline Wall loss $w$ & 5 dB \\
 \hline Loading factor $\beta$ & 0.8\\
 \hline ON/OFF algorithm thresholds $\epsilon_1,\epsilon_2$ & 0.1\\
 \hline  OFFLOAD parameter $\gamma$ & 1\\
  \hline
 \end{tabular}
 \label{table:sim-param}
\end{table}

\begin{figure}
  \centering
  \subfigure[$N_{\rm f} = 20$]{
  \includegraphics[width=0.47 \textwidth]{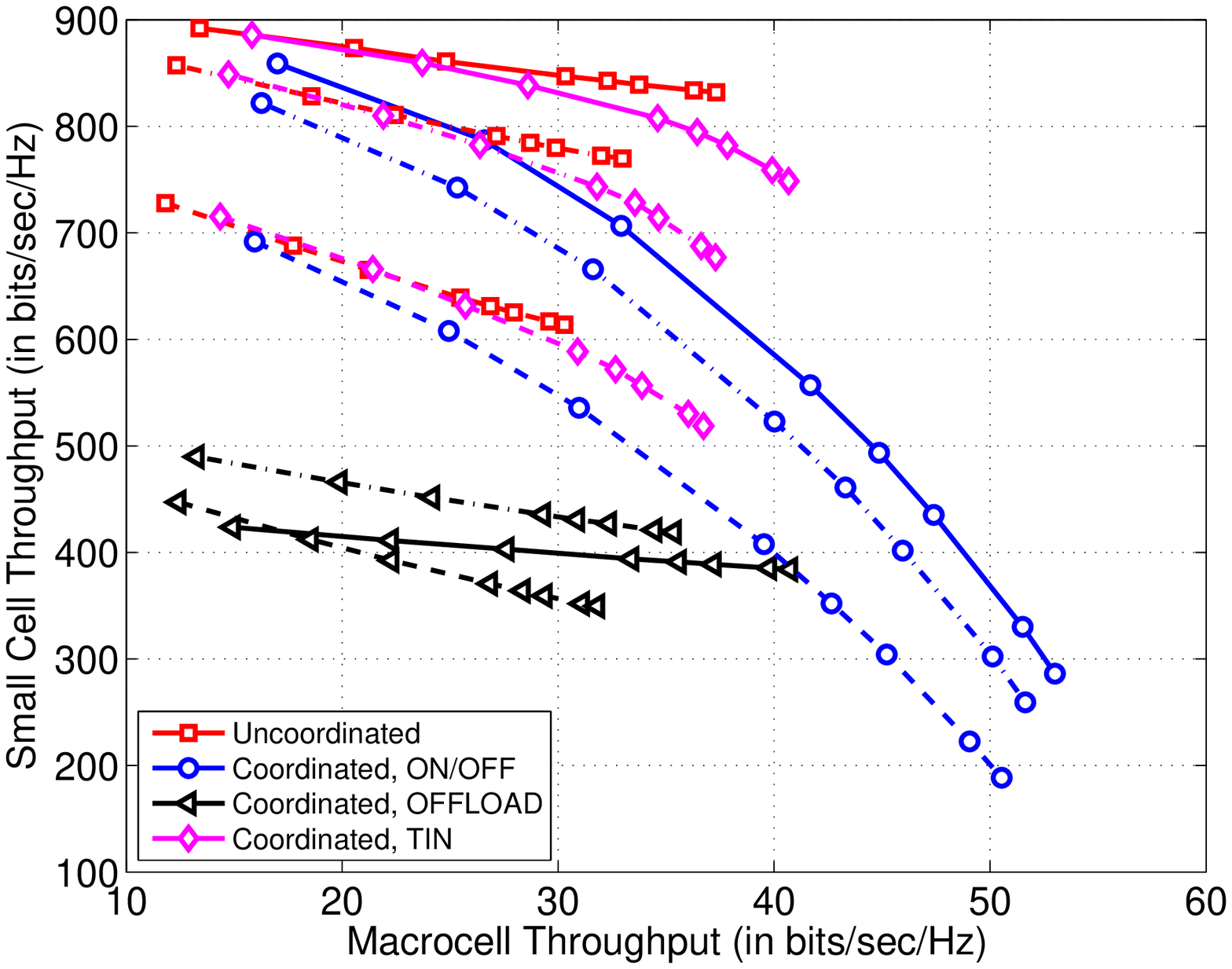} \label{fig:tradeoff-20}
  }
  \subfigure[$N_{\rm f} = 50$]{
  \includegraphics[width=0.47 \textwidth]{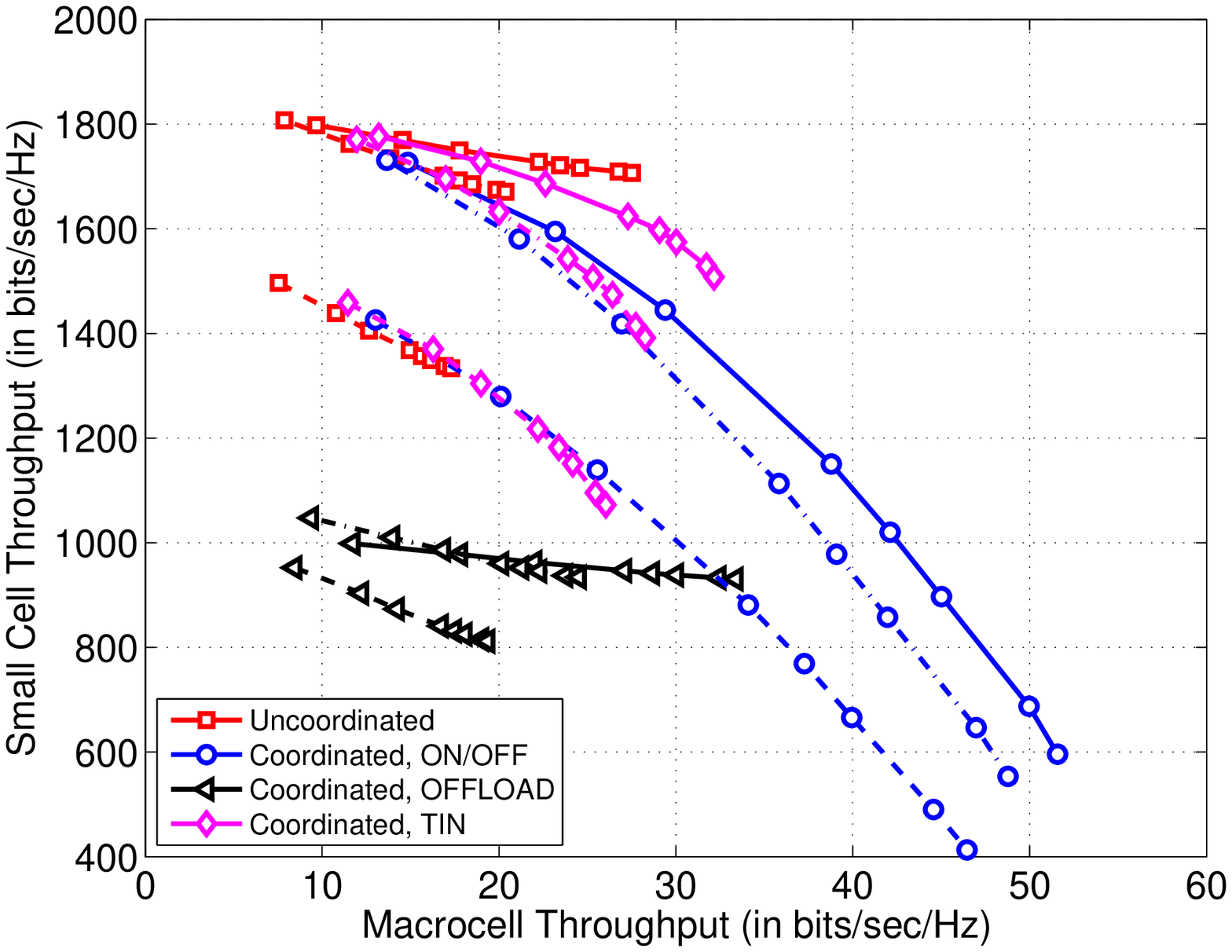} \label{fig:tradeoff-50}
  }
  \caption{Throughput tradeoff curves for different inter-tier coordination schemes and different deployments obtained by varying the number of user groups $G$ served by the macrocell.}\label{fig:tradeoff}
\end{figure}

Figure \ref{fig:tradeoff} shows the tradeoff between total macrocell throughput and total small cell throughput for different small cell deployments and various inter-tier interference coordination strategies discussed in detail in Section \ref{sec:inter-tier-ic}. The tradeoff curves are obtained by varying the parameter $G$, the number of user groups being served by the macrocell from 1 to 10, i.e., the leftmost marker in each plot corresponds to $G=1$ and the rightmost to $G=10$. We consider $N_{\rm f} = 20$ in Fig. \ref{fig:tradeoff-20} and $N_{\rm f} = 50$ in Fig. \ref{fig:tradeoff-50}. For the OFFLOAD coordination strategy of Section \ref{subsec:offload}, we consider the offloaded macrocell user groups as ``part'' of the small cell, which means their throughput is counted towards the small cell throughput. Increasing $N_{\rm f}$ results in increased small cell throughput but reduced macrocell throughput due to increased interference from small cells. We see that in order to increase the macrocell throughput at the expense of decreasing the small cell throughput, ON/OFF coordination strategy of Section \ref{subsec:on-off} gives good performance, followed by the coordination strategy involving TIN in Section \ref{subsec:tin}. Note that since the trade-off plots do not provide any information about the distribution of rates, they do not provide insights into strategies, such as the OFFLOAD strategy, which is specifically designed and optimized to ``equalize'' rates across different user groups. Nevertheless, even in terms of total throughput and for $\gamma=1$, OFFLOAD strategy provides a good tradeoff between macrocell and small cell rates, especially when $G$ is high. We will discuss this strategy and study its performance as a function of $\gamma$ later in this section. Overall, an interesting observation that applies to almost all the setups is that the system performs better when small cells are deployed at the edge of the macrocell. This is because it simultaneously reduces interference to the user groups served by macrocell and small cells. For the small cells that are more likely to be at the edge, the macrocell interference is weak at those user groups due to higher path-loss. Also, these small cells do not interfere with the cell-interior macrocell user groups, again due to higher path loss, leading to higher rates in general.

\begin{figure}
  \centering
  \subfigure[$N_{\rm f} = 20$]{
  \includegraphics[width=0.48 \textwidth]{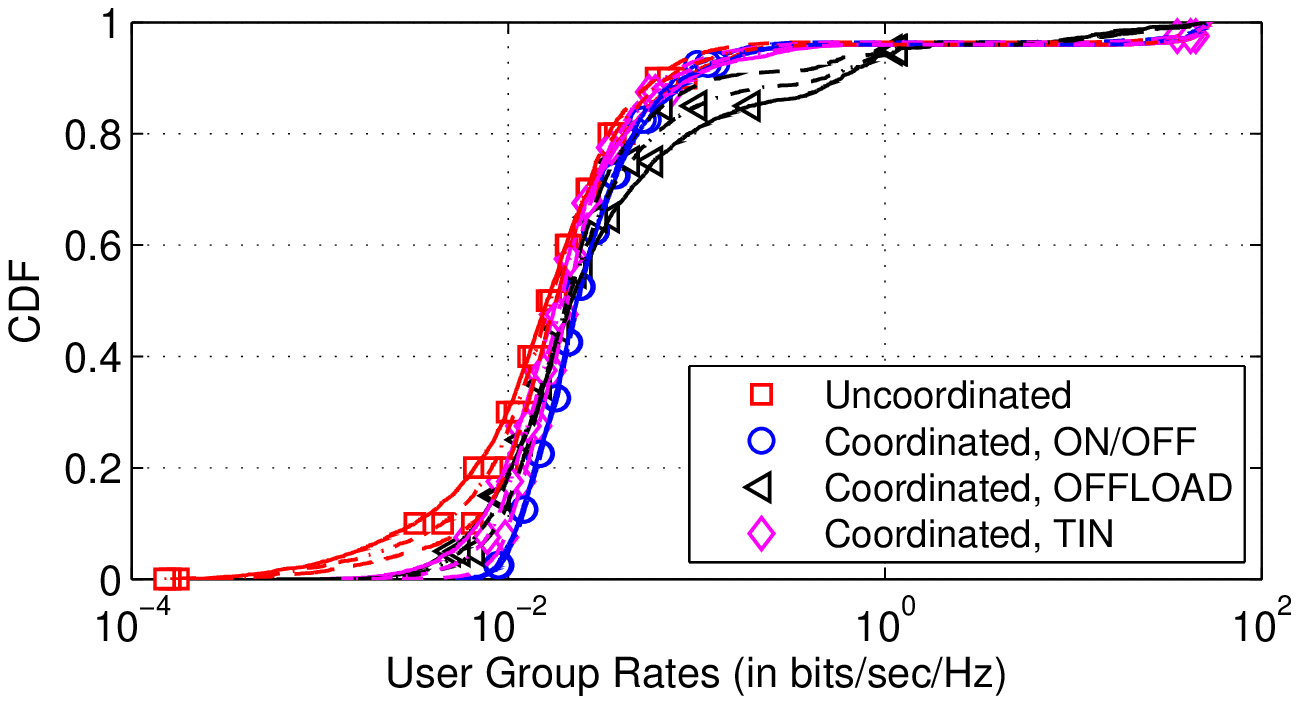}}
  \subfigure[$N_{\rm f} = 50$]{
  \includegraphics[width=0.48 \textwidth]{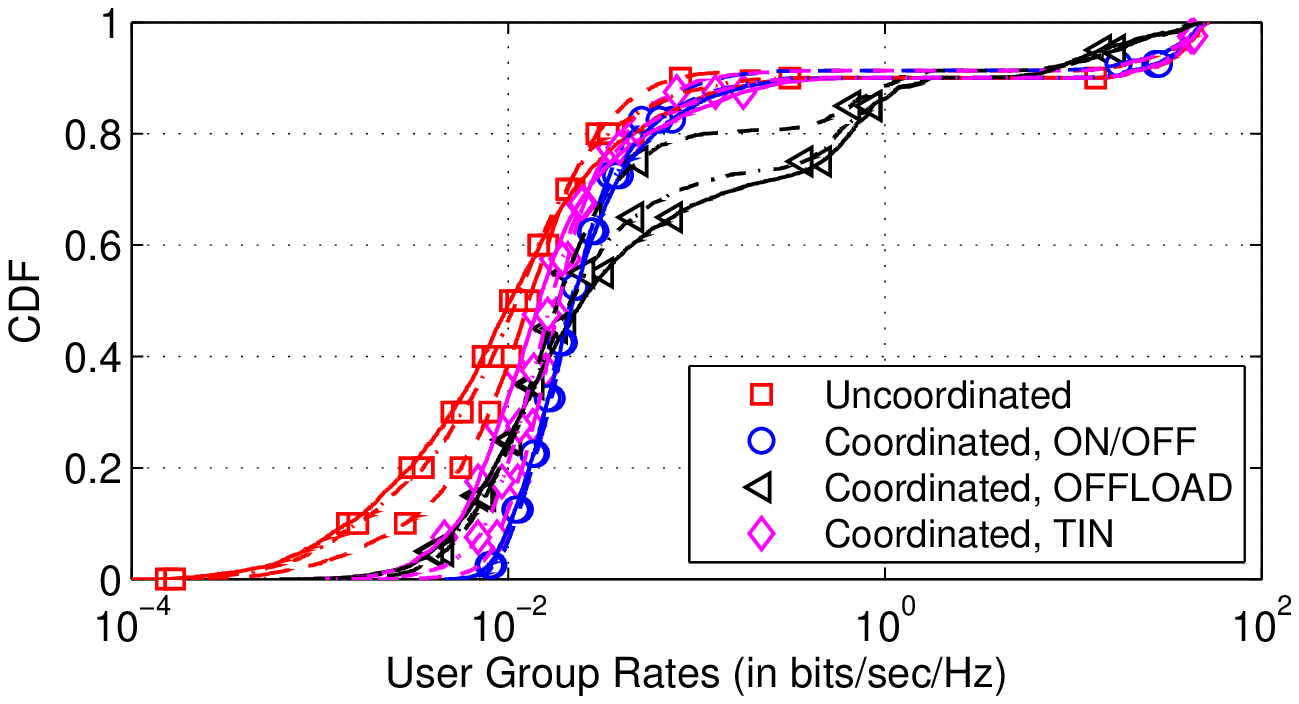}}
  \caption{CDF of user group rates for different inter-tier coordination schemes and different deployments ($G = 1$).}\label{fig:G-1-cdf}
\end{figure}

%\begin{figure}
%  \centering
%  \subfigure[$N_{\rm f} = 20$]{
%  \includegraphics[width=0.48 \textwidth]{rates_all_20_G_5}}
%  \subfigure[$N_{\rm f} = 50$]{
%  \includegraphics[width=0.48 \textwidth]{rates_all_50_G_5}}
%  \caption{CDF of user group rates for different inter-tier coordination schemes and different deployments ($G = 5$). The `solid' lines denote the cell edge deployment, the `dashed' lines denote the cell interior deployment and the `dash dotted' lines denote the scenario for uniform deployment of small cells.}\label{fig:G-5-cdf}
%\end{figure}

\begin{figure}
  \centering
  \subfigure[$N_{\rm f} = 20$]{
  \includegraphics[width=0.48 \textwidth]{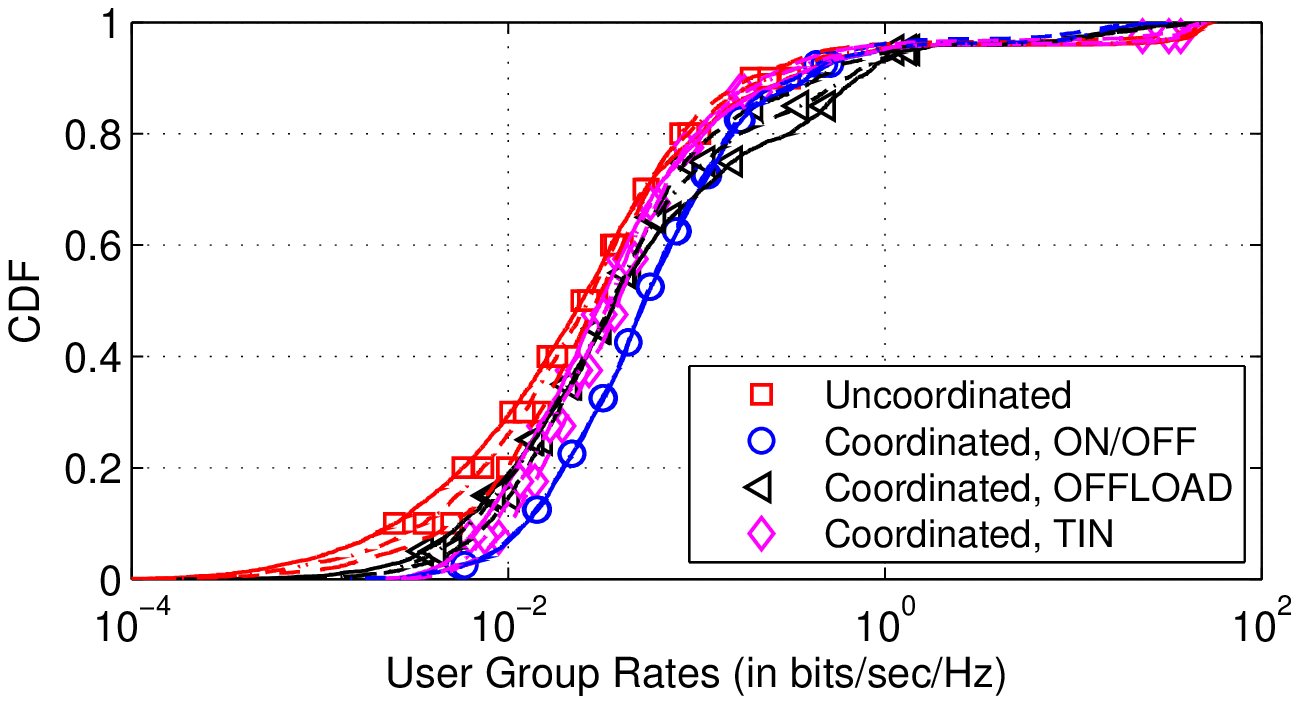}}
  \subfigure[$N_{\rm f} = 50$]{
  \includegraphics[width=0.48 \textwidth]{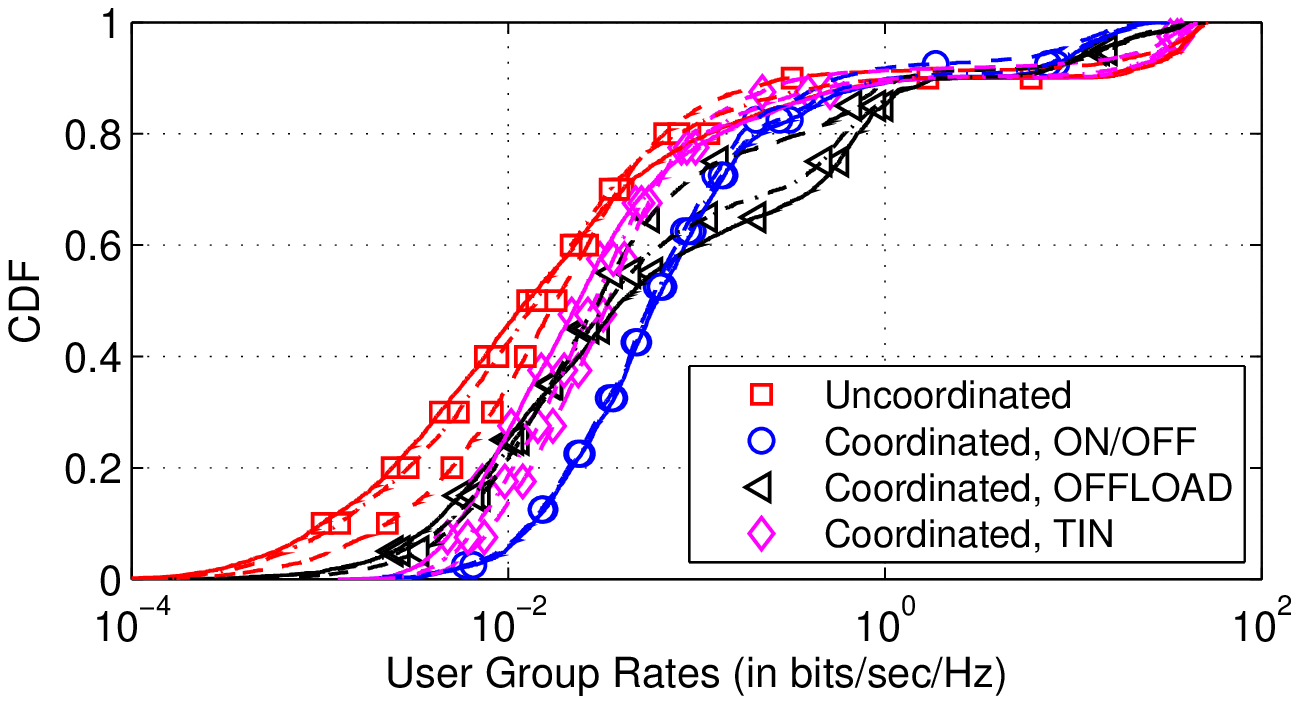}}
  \caption{CDF of user group rates for different inter-tier coordination schemes and different deployments ($G = 10$).}\label{fig:G-10-cdf}
\end{figure}

Figures \ref{fig:G-1-cdf} and \ref{fig:G-10-cdf} show the CDFs of the user group rates for different small cell deployments and various coordination strategies, when the macrocell serves $G = 1$ and $10$ user groups, respectively. In each of the figures, we show the rate CDFs for $N_{\rm f} = 20$ and $N_{\rm f} = 50$. The low rates are the rates observed at the macrocell user groups and the high rates are the rates at the user groups served by small cells. It can be seen from the plots that the uncoordinated strategy performs the worst when it comes to the macrocell user group rates, because of the increased cross-tier interference due to all the transmitting small cells. In general, the macrocell user group rates show a decreasing trend on going from a cell interior deployment to a cell edge deployment of the small cells, because of the reduced signal strength to the macrocell user groups located far from the macrocell. The ON/OFF strategy is the best in terms of guaranteeing a good macrocell throughput, albeit at the cost of decreased small cell throughput due to the fact that some of the small cells shut down their transmissions in every slot. The TIN coordination strategy also increases the macrocell user group rates while causing a minor degradation in the rates of the user groups served by the small cells. The sharp transition in the plots at the higher end of the user group rates is because of the drastic difference in the rates observed at user groups served by the macrocell and those served by the small cell. The length of this transition (or the difference in the rates of the macrocell and small cell user groups) decreases when more number of user groups are served by the macrocell, which causes more interference to the small cells, thereby decreasing their throughput. The benefit of the OFFLOAD coordination strategy can be seen as increasing the rates of the macrocell user groups which are now being served by the small cell, which is prominent when more small cells are deployed at the edge. This is because of the fact that a user group located at the cell edge has a greater chance of being offloaded from the macrocell to a small cell because it receives a stronger signal from the latter. The disadvantage comes at the cost of reducing the rates of the user groups being originally served by the small cells, due to a sharing of transmission resources between its own user group and the offloaded user group. %In all the figures corresponding to the OFFLOAD coordination strategy, we set $\gamma = 1$.

\begin{figure}
  \centering
  \subfigure[$N_{\rm f} = 20$]{
  \includegraphics[width=0.48 \textwidth]{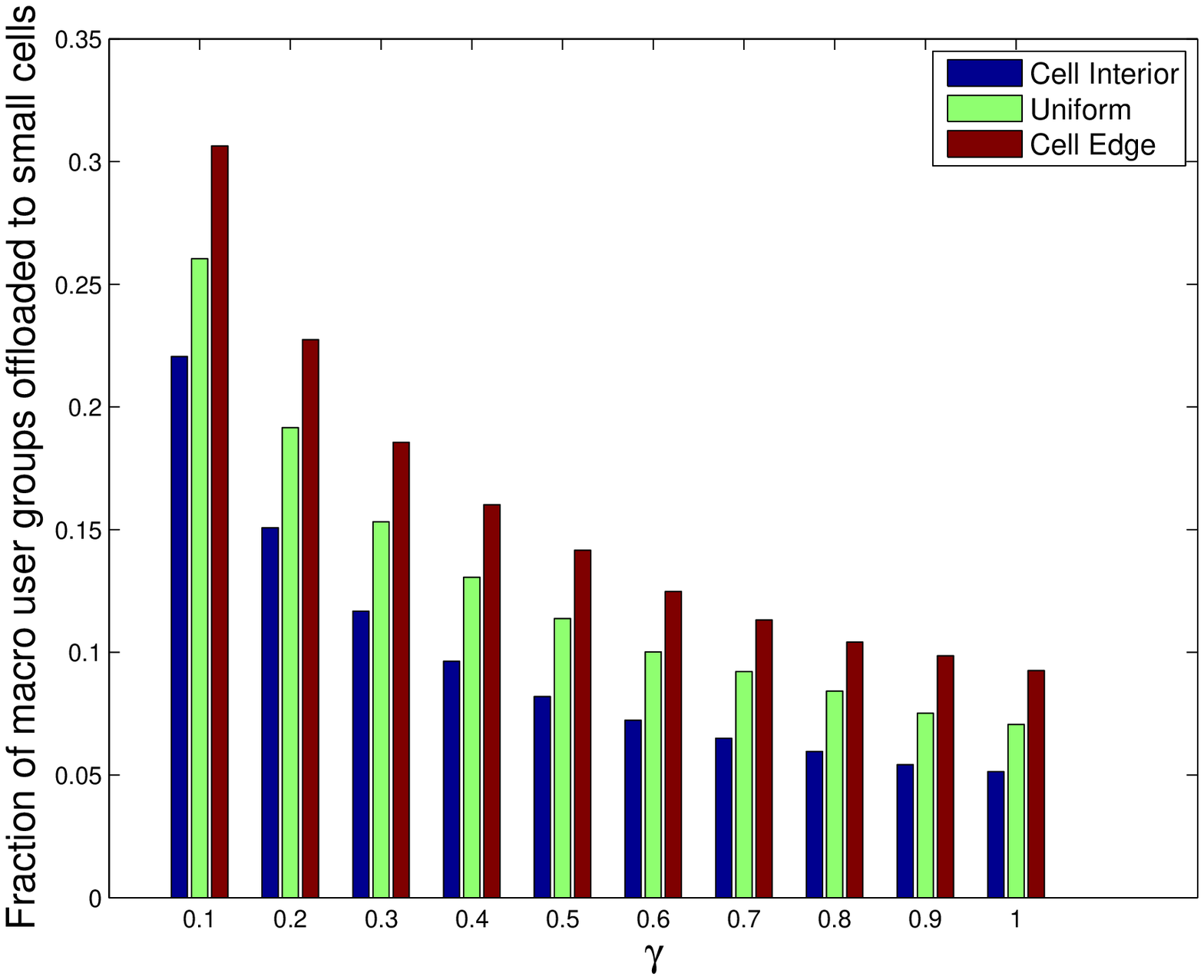}}
  \subfigure[$N_{\rm f} = 50$]{
  \includegraphics[width=0.48 \textwidth]{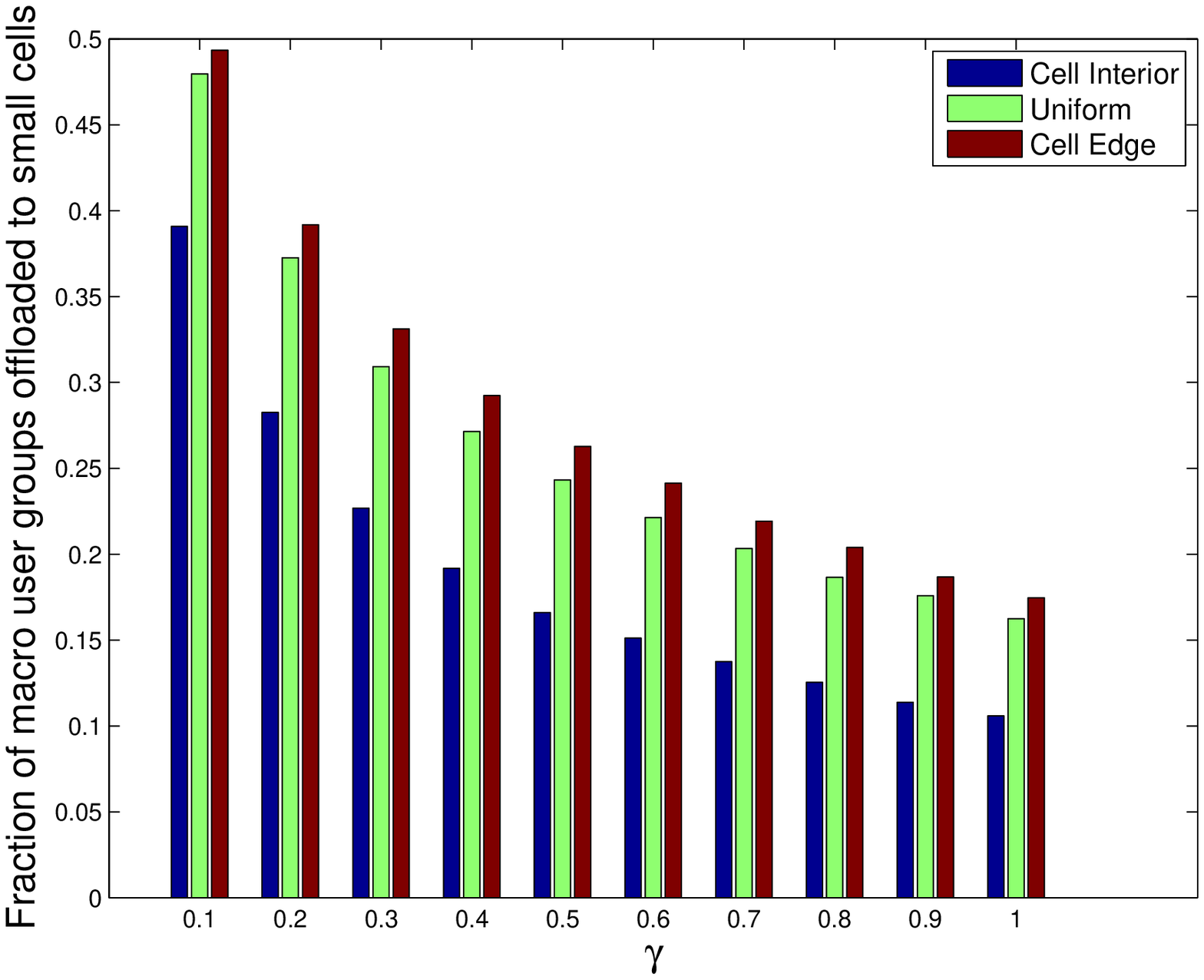}}
  \caption{Fraction of users offloaded from the macrocell to the small cells for different deployments and varying $\gamma$.}\label{fig:offload_frac}
\end{figure}

In all the results so far, we did not optimize the value of $\gamma$ for the OFFLOAD strategy, which we do now. Fig. \ref{fig:offload_frac} shows the fraction of offloaded users with varying $\gamma$, for various small cell deployments and different small cell densities, for the coordination scheme of Section \ref{subsec:offload}. The fraction of users being offloaded increases as $N_{\rm f}$ increases from 20 to 50, and decreases with an increase in the parameter $\gamma$. Thus, $\gamma$ controls the fraction of offloaded users. By having a larger value of $\gamma$, more preference is given to the signal strength from the macrocell, leading to less aggressive offloading. For a smaller $\gamma$, the small cells get higher priority, hence, the fraction of offloaded users is more. Also, the fraction of offloaded users increases as we go from a cell interior deployment to a cell edge deployment of the small cells, because of the decrease in the signal strength of the macrocell users as we go towards the cell edge.

%\begin{figure}
%  \centering
%  \subfigure[$N_{\rm f} = 20$]{
%  \includegraphics[width=0.48 \textwidth]{offload_20_G_1}}
%  \subfigure[$N_{\rm f} = 50$]{
%  \includegraphics[width=0.48 \textwidth]{offload_50_G_1}}
%  \caption{CDF of user group rates for coordination scheme `OFFLOAD' and different deployments ($G = 1$).}\label{fig:G-1-cdf-offload}
%\end{figure}

\begin{figure}
  \centering
  \subfigure[$N_{\rm f} = 20$]{
  \includegraphics[width=0.48 \textwidth]{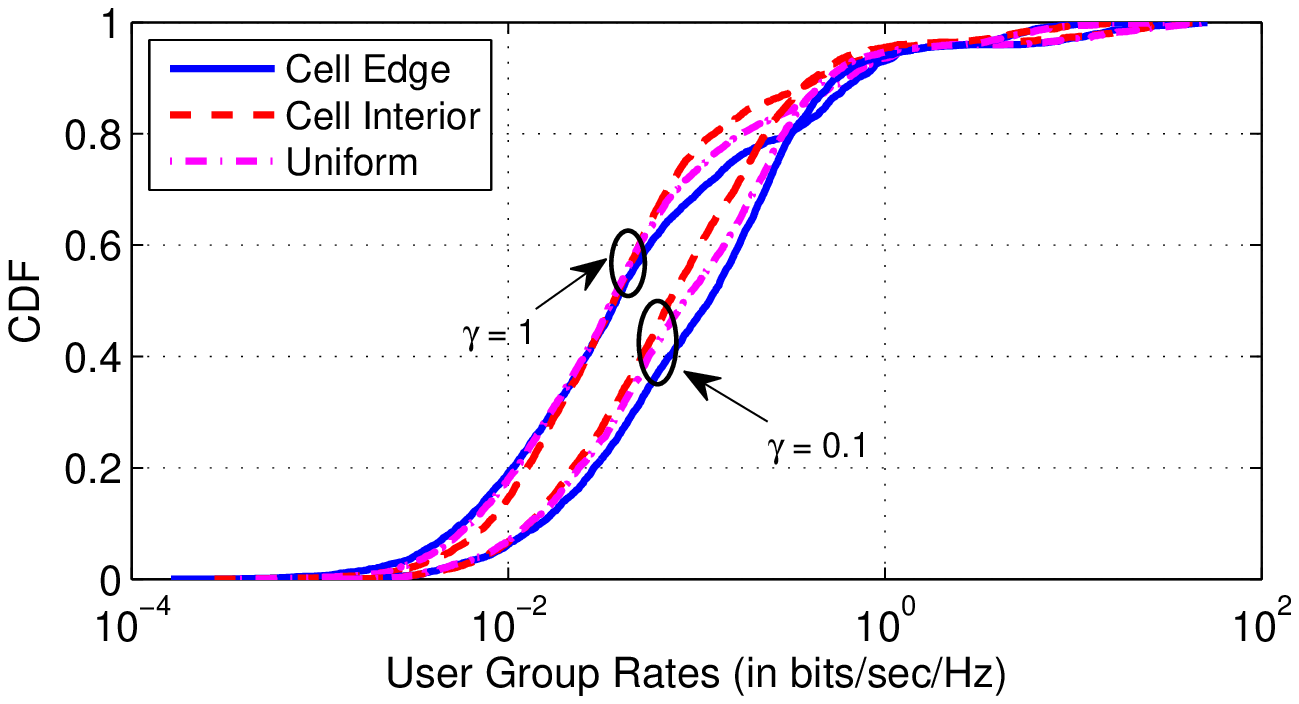}}
  \subfigure[$N_{\rm f} = 50$]{
  \includegraphics[width=0.48 \textwidth]{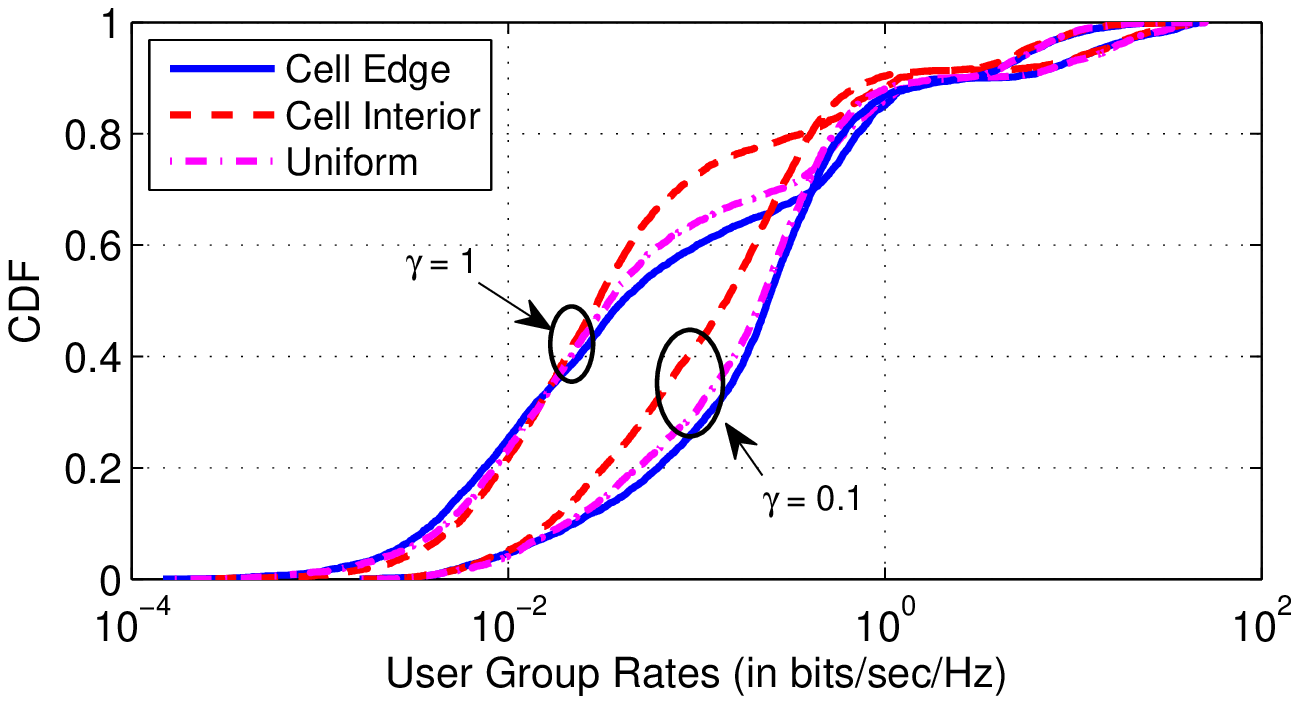}}
  \caption{CDF of user group rates for coordination scheme `OFFLOAD' and different deployments ($G = 10$).}\label{fig:G-10-cdf-offload}
\end{figure}

Figure \ref{fig:G-10-cdf-offload} shows the CDFs of the user group rates for the OFFLOAD coordination strategy for different small cell deployments and varying $\gamma$, when the macrocell serves $G = 10$ user groups. We see that a lower value of $\gamma$ favors offloading, thereby increasing the rates of the offloaded macrocell user groups resulting in an increase of the macrocell user group rates, although at the cost of reduced small cell rates. Note that these reduced small cell rates may not still be the system bottleneck due to limited backhaul capacity of the small cells. Therefore, aggressive offloading might be desirable to boost the macrocell rates while still keeping the small cell rates higher than the backhaul bottleneck.

\section{Conclusions}
In this paper, we considered a heterogeneous cellular network with the following features: (i) macrocells and small cells share the same spectrum, hence interfere with each other, (ii) users form hotspots (referred to as user groups in the paper), i.e., they are concentrated at certain areas in the cell, (iii) the size of a hotspot is much smaller than the macrocell radius, as a result of which the users at a given hotspot appear co-located to the macrocell, and (iv) some of the hotspots have a dedicated small cell in their vicinity while the rest have to be served by the macrocell. We further assume a large number of antennas at the macrocell (massive-MIMO), using which it can concentrate its transmission energy in the direction of hotspots it serves. This provides transmission opportunities to the small cells located in other directions (termed {\em spatial blanking} of macrocell). In addition to this implicit interference mitigation, we develop three low-complexity strategies for explicit inter-tier interference coordination. While the two strategies involve turning OFF small cells intelligently, the last one offloads macrocell traffic to small cells thereby providing significant throughput gains. Our analysis also provides insights into where exactly the deployment of small cells provides most benefits for a given performance metric, e.g., uniform vs. cell-interior vs. cell-edge. Note that cell-edge is not an {\em a priori} intuitive choice in this case due to directional channels and spatial blanking.

A straightforward extension of this work includes considering the effect of having hotspots of different sizes on the downlink performance, which can be modeled by tuning the scattering radius. Further, in this work we ignored the presence of {\em isolated users}, i.e., users that are not a part of any hotspot. A concrete direction of future work would include these users and study their effect on the throughput of the hotspot users (note that the spatial resources will need to be shared by isolated users and the hotspots). Two other extensions include coordination across multiple macrocells and a similar analysis as this paper for the cellular uplink.

\appendices

\section{TIN Selection Algorithm} \label{appendix-tin-algo}

We outline the steps of ``TIN selection algorithm'' for choosing a set of user groups that satisfy the TIN optimality condition in the resulting network. Recall that all the user data streams sent by the macrocell are transmitted with equal power $\frac{P_0}{S}$ and those sent by the small cells are transmitted with power $\frac{P_1}{\bar{S}}$.
The direct link gain of users in group $g \in \Mc$ served by the macrocell is given by $D_{g,0}^{\rm mc}$ and that of the users in group $f \in \Sc$
served by the corresponding small cell is given by $D_{f,f}^{\rm sc}$. Similarly, we have $I^{\rm mc}_{g,g'}$, $J^{\rm sc}_{g,f}$, $I^{\rm sc}_{f,f'}$ and $J^{\rm mc}_{f,g}$ for the cross channel  gains, where $I^{\rm mc}_{g,g'}$ denotes the inter-group interference between user groups $g$ and $g'$ in $\Mc$, $I^{\rm sc}_{f,f'}$ denotes the interference between user groups $f$ and $f'$ in $\Sc$, $J^{\rm mc}_{f,g}$ is the cross-tier interference caused by the data streams sent by the macrocell to group $g \in \Mc$ on users in group $f \in \Sc$, and $J^{\rm sc}_{g,f}$ denotes the inter-tier interference caused small cell $f \in \Sc$ on users of group $g \in \Mc$. Expressions for $D_{g,0}^{\rm mc}$, $D_{f,f}^{\rm sc}$, $I^{\rm mc}_{g,g'}$, $J^{\rm sc}_{g,f}$, $I^{\rm sc}_{f,f'}$ and $J^{\rm mc}_{f,g}$ are given in Section \ref{sec:da-sinr}. In addition, $\cv$ denotes the vector of user group priorities that is updated at every scheduling slot in order to guarantee equal air-time to all the users in the network.
\begin{itemize}
\item {\bf Step 1:} We start by forming a set of user groups ($\Cc_{\rm mc}$ for the macrocell user groups and $\Cc_{\rm sc}$ for the small cell user groups) that have the highest priority. From this set, we select the user group that has the highest direct link gain. Let $c_{\max}$ be the maximum element of the vector $\cv$. Define two sets $\Cc_{\rm mc} = \emptyset$ and $\Cc_{\rm sc} = \emptyset$. For every $g \in \Mc$ and $f \in \Sc$, update
\begin{eqnarray}
\Cc_{\rm mc} = \Cc_{\rm mc} \bigcup g \ \  {\rm if} \ \ c_{g} = c_{\max},\ && \Cc_{\rm sc} \bigcup f \ \  {\rm if} \ \ c_{f} = c_{\max} \nonumber
\end{eqnarray}
Find $g^*$ and $f^*$ such that
\begin{eqnarray}
g^* = {{\rm arg} \max}_{g \in \Cc_{\rm mc}} D_{g,0}^{\rm mc} \frac{P_0}{S_g},&& f^* = {{\rm arg} \max}_{f \in \Cc_{\rm sc}} D_{f,f}^{\rm sc} \frac{P_1}{\bar{S}}
\end{eqnarray}
If $D_{g,0}^{\rm mc} \frac{P_0}{S_g} > D_{f,f}^{\rm sc} \frac{P_1}{\bar{S}}$, initialize $\Gc = g^*$, $\Sc_A = \emptyset$, $\Mc^{(0)}_{\rm res} = \Mc \setminus g^*$, $\Sc^{(0)}_{\rm res} = \Sc$, $S^{(0)} = \sum_{g \in \Gc} = S_{g^*}$. Else, initialize $\Gc = \emptyset$, $\Sc_A = f^*$, $\Mc^{(0)}_{\rm res} = \Mc$, $\Sc^{(0)}_{\rm res} = \Sc \setminus f^*$, $S^{(0)} = \sum_{g \in \Gc} = 0$. Note that $\Gc$ contains the set of selected macrocell user groups, $\Sc_A$ the set of selected small cell user groups and $\Mc_{\rm res}^{(n)}$ and $\Sc_{\rm res}^{(n)}$ the set of macrocell and small cell user groups that have not been selected at iteration $n$ respectively.
\item {\bf Step 2:}  At every iteration $n$ of the algorithm, we find those user groups in $\Mc_{\rm res}^{(n)}$ and $\Sc_{\rm res}^{(n)}$ that can be added to $\Gc$ and $\Sc_A$ without violating the TIN optimality conditions (\ref{eqn:tin-opt-condn}) and store them in $\Gc_{\rm TIN}$ and $\Sc_{\rm TIN}$. In order to do this, $\forall \ g \in \Mc^{(n)}_{\rm res}$, we define the following three quantities
\begin{eqnarray}
\kappa^g_{g',{\rm mc}} &=& \frac{D_{g',0}^{\rm mc} \frac{P_0}{S^{(n)} + S_g}}{\left[\max\left(C^{\rm mc}_{1,g',g},\frac{P_0}{S^{(n)} + S_g} I^{\rm mc}_{g',g} \right) \right] \times \left[ \max \left( C^{\rm mc}_{2,g',g}, \frac{P_0}{S^{(n)} + S_g} I^{\rm mc}_{g,g'} \right) \right]} \ \forall \ g' \in \Gc \nonumber\\
\kappa^g_{f,{\rm sc}} &=& \frac{D_{f,f}^{\rm sc} \frac{P_1}{\bar{S}} }{\left[\max\left(C^{\rm sc}_{1,f,g},J^{\rm mc}_{f,g} \frac{P_0}{S^{(n)} + S_g} \right) \right] \times \left[ \max \left( C^{\rm sc}_{2,f,g}, I^{\rm sc}_{g,f} \frac{P_1}{\bar{S}} \right) \right]} \ \forall \ f \in \Sc_A \nonumber\\
\kappa^g_{\rm self,mc} &=& \frac{D_{g,0}^{\rm mc} \frac{P_0}{S^{(n)} + S_g}}{\left[C^{\rm self,mc}_{1,g}\right] \times \left[C^{\rm self,mc}_{2,g}\right]} \nonumber
\end{eqnarray}
where
\begin{eqnarray}
C^{\rm mc}_{1,g',g} &=& \max \left[ \frac{P_0}{S^{(n)} + S_g} \max_{m \in \Gc, m \neq g'} I^{\rm mc}_{g',m} , \frac{P_1}{\bar{S}} \max_{f \in \Sc_A} J^{\rm sc}_{g',f} \right] \nonumber\\
C^{\rm mc}_{2,g',g} &=& \max \left[ \frac{P_0}{S^{(n)} + S_g} \max_{m \in \Gc, m \neq g'} I^{\rm mc}_{m,g'} , \frac{P_0}{S^{(n)} + S_g} \max_{f \in \Sc_A} J^{\rm mc}_{f,g'} \right] \nonumber\\
C^{\rm sc}_{1,f,g} &=& \max \left[ \frac{P_0}{S^{(n)} + S_g} \max_{m \in \Gc} J^{\rm mc}_{f,m} , \frac{P_1}{\bar{S}} \max_{f' \in \Sc_A, f' \neq f} I^{\rm sc}_{f,f'} \right] \nonumber\\
C^{\rm sc}_{2,f,g} &=& \max \frac{P_1}{\bar{S}} \left[ \max_{m \in \Gc} J^{\rm sc}_{m,f} , \max_{f' \in \Sc_A, f' \neq f} I^{\rm sc}_{f',f} \right] \nonumber\\
C^{\rm self,mc}_{1,g} &=& \max \left[ \frac{P_0}{S^{(n)} + S_g} \max_{m \in \Gc} I^{\rm mc}_{g,m} , \frac{P_1}{\bar{S}} \max_{f \in \Sc_A} J^{\rm sc}_{g,f} \right] \nonumber\\
C^{\rm self,mc}_{2,g} &=& \max \left[ \frac{P_0}{S^{(n)} + S_g} \max_{m \in \Gc} I^{\rm mc}_{m,g} , \frac{P_0}{S^{(n)} + S_g} \max_{f \in \Sc_A} J^{\rm mc}_{f,g} \right] \nonumber
\end{eqnarray}
Set $\Gc_{\rm TIN} = \emptyset$ and $\forall \ g \in \Mc^{(n)}_{\rm res}$, make the assignment
$\Gc_{\rm TIN} = \Gc_{\rm TIN} \bigcup g$ if the following conditions are satisfied for user group $g$
\begin{eqnarray}
\kappa^g_{g',{\rm mc}} > 1 \ \forall \ g' \in \Gc,\ \ \ \kappa^g_{f,{\rm sc}} > 1 \ \forall \ f \in \Sc_A ,\ \ \ \kappa^g_{{\rm self,mc}} > 1 \nonumber
\end{eqnarray}
which are essentially the TIN optimality conditions that a macrocell user group $g$ needs to satisfy.
\item {\bf Step 3:} Similarly, $\forall \ f \in \Sc^{(n)}_{\rm res}$, we define the following three quantities
\begin{eqnarray}
\kappa^f_{g',{\rm mc}} &=& \frac{D_{g',0}^{\rm mc} \frac{P_0}{S^{(n)}}}{\left[\max\left(A^{\rm mc}_{1,g',f},\frac{P_1}{\bar{S}} J^{\rm sc}_{g',f} \right) \right] \times \left[ \max \left( A^{\rm mc}_{2,g',f}, \frac{P_0}{S^{(n)}} J^{\rm mc}_{f,g'} \right) \right]} \ \forall \ g' \in \Gc \nonumber\\
\kappa^f_{f',{\rm sc}} &=& \frac{D_{f',f'}^{\rm sc} \frac{P_1}{\bar{S}} }{\left[\max\left(A^{\rm sc}_{1,f',f},I^{\rm sc}_{f',f} \frac{P_1}{\bar{S}} \right) \right] \times \left[ \max \left( A^{\rm sc}_{2,f',f}, I^{\rm sc}_{f,f'} \frac{P_1}{\bar{S}} \right) \right]} \ \forall \ f' \in \Sc_A \nonumber\\
\kappa^f_{\rm self,sc} &=& \frac{D_{f,f}^{\rm sc} \frac{P_1}{\bar{S}}}{\left[A^{\rm self,sc}_{1,g}\right] \times \left[A^{\rm self,sc}_{2,g}\right]} \nonumber
\end{eqnarray}
where
\begin{eqnarray}
A^{\rm mc}_{1,g',f} &=& \max \left[ \frac{P_0}{S^{(n)}} \max_{m \in \Gc, m \neq g'} I^{\rm mc}_{g',m} , \frac{P_1}{\bar{S}} \max_{f \in \Sc_A} J^{\rm sc}_{g',f} \right] \nonumber\\
A^{\rm mc}_{2,g',f} &=& \max \left[ \frac{P_0}{S^{(n)}} \max_{m \in \Gc, m \neq g'} I^{\rm mc}_{m,g'} , \frac{P_0}{S^{(n)}} \max_{f \in \Sc_A} J^{\rm mc}_{f,g'} \right] \nonumber\\
A^{\rm sc}_{1,f',f} &=& \max \left[ \frac{P_0}{S^{(n)}} \max_{m \in \Gc} J^{\rm mc}_{f',m} , \frac{P_1}{\bar{S}} \max_{f'' \in \Sc_A, f'' \neq f'} I^{\rm sc}_{f',f''} \right] \nonumber\\
A^{\rm sc}_{2,f',f} &=& \max \frac{P_1}{\bar{S}} \left[ \max_{m \in \Gc} J^{\rm sc}_{m,f'} , \max_{f'' \in \Sc_A, f'' \neq f'} I^{\rm sc}_{f'',f'} \right] \nonumber\\
A^{\rm self,sc}_{1,g} &=& \max \left[ \frac{P_0}{S^{(n)}} \max_{m \in \Gc} J^{\rm mc}_{f,m} , \frac{P_1}{\bar{S}} \max_{f' \in \Sc_A} I^{\rm sc}_{f,f'} \right] \nonumber\\
A^{\rm self,sc}_{2,g} &=& \max \left[ \frac{P_1}{\bar{S}} \max_{m \in \Gc} J^{\rm sc}_{m,f} , \frac{P_1}{\bar{S}} \max_{f' \in \Sc_A} I^{\rm mc}_{f',f} \right] \nonumber
\end{eqnarray}
Set $\Sc_{\rm TIN} = \emptyset$ and $\forall \ g \in \Sc^{(n)}_{\rm res}$, make the assignment $\Sc_{\rm TIN} = \Sc_{\rm TIN} \bigcup f$ if the following conditions are satisfied for user group $f$
\begin{eqnarray}
\kappa^f_{g',{\rm mc}} > 1 \ \forall \ g' \in \Gc, \ \ \ \kappa^f_{f',{\rm sc}} > 1 \ \forall \ f' \in \Sc_A, \ \ \ \kappa^f_{{\rm self,sc}} > 1 \nonumber
\end{eqnarray}
Note that these are the TIN optimality conditions that a small cell user group $f$ needs to satisfy to be added to $\Sc_{\rm TIN}$.
\item {\bf Step 4:} If there are no user groups that can be added without violating the TIN optimality conditions, we terminate the algorithm. Precisely, if $\Gc_{\rm TIN} = \emptyset$ and $\Fc_{\rm TIN} = \emptyset$, go to Step 8. Otherwise, we form a set of user groups from $\Gc_{\rm TIN}$ and $\Sc_{\rm TIN}$ which have the highest priority, similar to Step 1 and from this set, select a user group according to a heuristic given by the product of the $\kappa$ terms in Steps 2 and 3. Note that $\kappa$ is a ratio of the direct link signal strength to the strength of the interfering links, implying that a higher value of $\kappa$ means a more favorable link. We use a product of these terms for the already selected user groups and the user group in consideration and choose the user group with the maximum value of the product. In order to do this, we let $c_{\max}$ be the maximum element of the vector $\cv$ and define two sets $\Cc_{\rm mc} = \emptyset$ and $\Cc_{\rm sc} = \emptyset$. For every $g \in \Gc_{\rm TIN}$ and $f \in \Sc_{\rm TIN}$, we have
\begin{eqnarray}
\Cc_{\rm mc} = \Cc_{\rm mc} \bigcup g \ \  {\rm if} \ \ c_{g} = c_{\max},&& \Cc_{\rm sc} = \Cc_{\rm sc} \bigcup f \ \  {\rm if} \ \ c_{f} = c_{\max} \nonumber
\end{eqnarray}
 \item {\bf Step 5:} Find $g^*$ and $f^*$ such that
 $$g^* = {{\rm arg} \max}_{g \in \Cc_{\rm mc}} \left[ \prod_{g' \in \Gc} \kappa^g_{g',{\rm mc}} \times \prod_{f \in \Sc_A}  \kappa^g_{f,{\rm sc}} \times \kappa^g_{\rm self,mc} \right]$$
 $$f^* = {{\rm arg} \max}_{f \in \Cc_{\rm sc}} \left[ \prod_{g' \in \Gc} \kappa^f_{g',{\rm mc}} \times \prod_{f' \in \Sc_A} \kappa^f_{f',{\rm sc}} \times \kappa^f_{\rm self,sc} \right]$$
 \item {\bf Step 6:} If $\left[ \prod_{g' \in \Gc}  \kappa^{g*}_{g',{\rm mc}} \times \prod_{f \in \Sc_A} \kappa^{g*}_{f,{\rm sc}} \times \kappa^{g*}_{\rm self,mc} \right] > \left[ \prod_{g' \in \Gc} \kappa^{f*}_{g',{\rm mc}} \times \prod_{f' \in \Sc_A} \kappa^{f*}_{f',{\rm sc}} \times \kappa^{f*}_{\rm self,sc} \right]$, update
 $$\Gc = \Gc \bigcup g^*, \ \ S^{(n+1)} = \sum_{g \in \Gc} S_g, \ \ \Mc_{\rm res}^{(n+1)} = \Mc_{\rm res}^{(n)} \setminus g^*, \ \ \Sc_{\rm res}^{(n+1)} = \Sc_{\rm res}^{(n)}$$
 else
 $$\Sc_A = \Sc_A \bigcup f^*, \ \ S^{(n+1)} = S^{(n)}, \ \ \Mc_{\rm res}^{(n+1)} = \Mc_{\rm res}^{(n)}, \ \ \Sc_{\rm res}^{(n+1)} = \Sc_{\rm res}^{(n)} \setminus f^*$$
 Note that after user group selection, the sets $\Mc_{\rm res}^{(n)}$ and $\Sc_{\rm res}^{(n)}$ are updated accordingly.
 \item {\bf Step 7:} We check whether more macrocell user groups can be added and go to the corresponding step after incrementing the iteration, i.e., if $|\Gc| = G$, $\Fc_{\rm TIN} \neq \emptyset$ or $|\Gc| < G$, $\Gc_{\rm TIN} = \emptyset$, $\Fc_{\rm TIN} \neq \emptyset$, increment $n$ by 1 and go to Step 3. For all other cases, increment $n$ by 1 and go to Step 2.
 \item {\bf Step 8:} Output $\Gc$ and $\Sc_A$ as the result.

\end{itemize}
\linespread{1.3}
\bibliographystyle{IEEEtran}
\bibliography{references,Dhillon}

\end{document}

%% file: notation_HD.tex
\def\nb0{{\mathbf{0}}}
\def\nb1{{\mathbf{1}}}

\newtheorem{definition}{Definition}

\newtheorem{example}{Example}
\newtheorem{remark}{Remark}

%% file: macros.tex
\long\def\comment#1{}

\newfont{\bbb}{msbm10 scaled 700}

\newfont{\bbc}{msbm10 scaled 1100}
\newcommand{\CC}{\mbox{\bbc C}}

\newcommand{\cv}{{\pmb c}}
\newcommand{\dv}{{\pmb d}}

\newcommand{\hv}{{\pmb h}}

\newcommand{\pvx}{{\bf p}}
\newcommand{\qv}{{\pmb q}}

\newcommand{\sv}{{\pmb s}}

\newcommand{\wv}{{\pmb w}}

\newcommand{\yv}{{\pmb y}}
\newcommand{\zv}{{\pmb z}}

\newcommand{\Bm}{{\pmb B}}

\newcommand{\Hm}{{\pmb H}}
\newcommand{\Id}{{\pmb I}}

\newcommand{\Pm}{{\pmb P}}
\newcommand{\Qm}{{\pmb Q}}
\newcommand{\Rm}{{\pmb R}}

\newcommand{\Tm}{{\pmb T}}
\newcommand{\Um}{{\pmb U}}

\newcommand{\Cc}{{\cal C}}

\newcommand{\Fc}{{\cal F}}
\newcommand{\Gc}{{\cal G}}

\newcommand{\Ic}{{\cal I}}

\newcommand{\Mc}{{\cal M}}
\newcommand{\Nc}{{\cal N}}

\newcommand{\Sc}{{\cal S}}

\newcommand{\Lambdam}{\hbox{\boldmath$\Lambda$}}

\newcommand{\herm}{{\sf H}}

%% file: JSAC2014_InterferenceCoordination_ver14.bbl
\begin{thebibliography}{10}
\providecommand{\url}[1]{#1}
\csname url@samestyle\endcsname
\providecommand{\newblock}{\relax}
\providecommand{\bibinfo}[2]{#2}
\providecommand{\BIBentrySTDinterwordspacing}{\spaceskip=0pt\relax}
\providecommand{\BIBentryALTinterwordstretchfactor}{4}
\providecommand{\BIBentryALTinterwordspacing}{\spaceskip=\fontdimen2\font plus
\BIBentryALTinterwordstretchfactor\fontdimen3\font minus
  \fontdimen4\font\relax}
\providecommand{\BIBforeignlanguage}[2]{{%
\expandafter\ifx\csname l@#1\endcsname\relax
\typeout{** WARNING: IEEEtran.bst: No hyphenation pattern has been}%
\typeout{** loaded for the language `#1'. Using the pattern for}%
\typeout{** the default language instead.}%
\else
\language=\csname l@#1\endcsname
\fi
#2}}
\providecommand{\BIBdecl}{\relax}
\BIBdecl

\bibitem{AdhDhiC2014}
A.~Adhikary, H.~S. Dhillon, and G.~Caire, ``Spatial blanking and inter-tier
  coordination in massive-{MIMO} heterogeneous cellular networks,'' submitted
  to {\em IEEE Globecom Workshops}, Austin, TX, Dec. 2014.

\bibitem{CisM2012}
Cisco, ``Cisco visual networking index: Global mobile data traffic forecast
  update, 2011 - 2016,'' white paper, Feb. 2012.

\bibitem{eicicperez}
D.~Lopez-Perez, I.~Guvenc, G.~D.~L. Roche, M.~Kountouris, T.~Quek, and
  J.~Zhang, ``Enhanced intercell interference coordination challenges in
  heterogenous networks,'' \emph{IEEE Wireless Commun.}, vol.~18, no.~3, pp. 22
  -- 30, Jun. 2011.

\bibitem{ghosh2012heterogeneous}
A.~Ghosh, N.~Mangalvedhe, R.~Ratasuk, B.~Mondal, M.~Cudak, E.~Visotsky, T.~A.
  Thomas, J.~G. Andrews, P.~Xia, H.~S. Jo \emph{et~al.}, ``Heterogeneous
  cellular networks: From theory to practice,'' \emph{IEEE Commun. Magazine},
  vol.~50, no.~6, pp. 54 -- 64, Jun. 2012.

\bibitem{BouPanJ2009}
G.~Boudreau, J.~Panicker, N.~Guo, R.~Chang, N.~Wang, and S.~Vrzic,
  ``Interference coordination and cancellation for 4{G} networks,'' \emph{IEEE
  Commun. Magazine}, vol.~47, no.~4, pp. 74 -- 81, Apr. 2009.

\bibitem{adhikary2011cognitive}
A.~Adhikary, V.~Ntranos, and G.~Caire, ``Cognitive femtocells: Breaking the
  spatial reuse barrier of cellular systems,'' \emph{Proc., Information Theory
  and its Applications (ITA)}, pp. 1 -- 10, Feb. 2011.

\bibitem{DhiGanJ2012}
H.~S. Dhillon, R.~K. Ganti, F.~Baccelli, and J.~G. Andrews, ``Modeling and
  analysis of {K}-tier downlink heterogeneous cellular networks,'' \emph{IEEE
  Journal on Sel. Areas in Commun.}, vol.~30, no.~3, pp. 550 -- 560, Apr. 2012.

\bibitem{NovGanJ2012}
T.~D. Novlan, R.~K. Ganti, A.~Ghosh, and J.~G. Andrews, ``Analytical evaluation
  of fractional frequency reuse for heterogeneous cellular networks,'' vol.~60,
  no.~7, pp. 2029 -- 2039, Jul. 2012.

\bibitem{SinAndJ2014}
S.~Singh and J.~G. Andrews, ``Joint resource partitioning and offloading in
  heterogeneous cellular networks,'' vol.~13, no.~2, pp. 888 -- 901, Feb. 2014.

\bibitem{CieWanJ2013}
M.~\v{C}ierny, H.~Wang, R.~Wichman, Z.~Ding, and C.~Wijting, ``On number of
  almost blank subframes in heterogeneous cellular networks,'' vol.~12, no.~10,
  pp. 5061 -- 5073, Oct. 2013.

\bibitem{DhiKouJ2013}
H.~S. Dhillon, M.~Kountouris, and J.~G. Andrews, ``Downlink {MIMO} {HetNets}:
  Modeling, ordering results and performance analysis,'' \emph{IEEE Trans. on
  Wireless Commun.}, vol.~12, no.~10, pp. 5208 -- 5222, Oct. 2013.

\bibitem{GupDhiJ2013}
A.~K. Gupta, H.~S. Dhillon, S.~Vishwanath, and J.~G. Andrews, ``Downlink
  multi-antenna heterogeneous cellular network with load balancing,'' submitted
  to {\em IEEE Trans. on Commun.}, Oct. 2013. Available online:
  arxiv.org/abs/1310.6795.

\bibitem{marzetta2010noncooperative}
T.~Marzetta, ``Noncooperative cellular wireless with unlimited numbers of base
  station antennas,'' \emph{IEEE Trans. on Wireless Commun.}, vol.~9, no.~11,
  pp. 3590 -- 3600, Nov. 2010.

\bibitem{wagner2012large}
S.~Wagner, R.~Couillet, M.~Debbah, and D.~T. Slock, ``Large system analysis of
  linear precoding in correlated {MISO} broadcast channels under limited
  feedback,'' \emph{IEEE Trans. on Info. Theory}, vol.~58, no.~7, pp. 4509 --
  4537, Jul. 2012.

\bibitem{adhikary2012joint}
A.~Adhikary, J.~Nam, J.-Y. Ahn, and G.~Caire, ``Joint spatial division and
  multiplexing: The large-scale array regime,'' \emph{IEEE Trans. on Info.
  Theory}, vol.~59, no.~10, pp. 6441 -- 6463, Oct. 2013.

\bibitem{adhikary2014massive}
A.~Adhikary, E.~A. Safadi, and G.~Caire, ``Massive {MIMO} and inter-tier
  interference coordination,'' \emph{Proc., Information Theory and its
  Applications (ITA)}, pp. 1 -- 10, Feb. 2014.

\bibitem{SinDhiJ2013}
S.~Singh, H.~S. Dhillon, and J.~G. Andrews, ``Offloading in heterogeneous
  networks: Modeling, analysis and design insights,'' \emph{IEEE Trans. on
  Wireless Commun.}, vol.~12, no.~5, pp. 2484 -- 2497, May 2013.

\bibitem{AndSinJ2014}
J.~G. Andrews, S.~Singh, Q.~Ye, X.~Lin, and H.~S. Dhillon, ``An overview of
  load balancing in {HetNets}: Old myths and open problems,'' \emph{IEEE
  Wireless Commun.}, vol.~21, no.~2, pp. 18 -- 25, Apr. 2014.

\bibitem{huh2012achieving}
H.~Huh, G.~Caire, H.~C. Papadopoulos, and S.~A. Ramprashad, ``Achieving
  ``massive {MIMO}'' spectral efficiency with a not-so-large number of
  antennas,'' \emph{IEEE Trans. on Wireless Commun.}, vol.~11, no.~9, pp. 3226
  -- 3239, Sep. 2012.

\bibitem{JunManJ2014}
V.~Jungnickel, K.~Manolakis, W.~Zirwas, B.~Panzner, V.~Braun, M.~Lossow,
  M.~Sternad, R.~Apelfrojd, and T.~Svensson, ``The role of small cells,
  coordinated multipoint, and massive {MIMO} in {5G},'' \emph{IEEE Commun.
  Magazine}, vol.~52, no.~5, pp. 44 -- 51, May 2014.

\bibitem{lee1973effects}
W.-Y. Lee, ``Effects on correlation between two mobile radio base-station
  antennas,'' \emph{IEEE Trans. on Veh. Technology}, vol.~22, no.~4, pp. 130 --
  140, Nov. 1973.

\bibitem{molisch2010wireless}
A.~F. Molisch, \emph{Wireless communications}.\hskip 1em plus 0.5em minus
  0.4em\relax Wiley. com, 2010.

\bibitem{adhikary2013joint2}
A.~Adhikary, E.~Al~Safadi, M.~Samimi, R.~Wang, G.~Caire, T.~S. Rappaport, and
  A.~F. Molisch, ``Joint spatial division and multiplexing for mm-wave
  channels,'' \emph{IEEE Journal on Sel. Areas in Commun., to appear}, 2014.

\bibitem{adhikary2013joint}
J.~Nam, A.~Adhikary, J.~Ahn, and G.~Caire, ``Joint spatial division and
  multiplexing: Opportunistic beamforming, user grouping and simplified
  downlink scheduling,'' \emph{IEEE Journal on Sel. Topics in Sig. Proc., to
  appear}, 2014.

\bibitem{yoo2006optimality}
T.~Yoo and A.~Goldsmith, ``On the optimality of multiantenna broadcast
  scheduling using zero-forcing beamforming,'' \emph{IEEE Journal on Sel. Areas
  in Commun.}, vol.~24, no.~3, pp. 528--541, Mar. 2006.

\bibitem{yoo2007multi}
T.~Yoo, N.~Jindal, and A.~Goldsmith, ``Multi-antenna downlink channels with
  limited feedback and user selection,'' \emph{IEEE Journal on Sel. Areas in
  Commun.}, vol.~25, no.~7, pp. 1478--1491, Sep. 2007.

\bibitem{dimic2005downlink}
G.~Dimic and N.~D. Sidiropoulos, ``On downlink beamforming with greedy user
  selection: performance analysis and a simple new algorithm,'' \emph{IEEE
  Trans. on Signal Processing}, vol.~53, no.~10, pp. 3857 -- 3868, Oct. 2005.

\bibitem{spencer2004zero}
Q.~H. Spencer, A.~L. Swindlehurst, and M.~Haardt, ``Zero-forcing methods for
  downlink spatial multiplexing in multiuser {MIMO} channels,'' \emph{IEEE
  Trans. on Signal Processing}, vol.~52, no.~2, pp. 461 -- 471, Feb. 2004.

\bibitem{shirani2010mimo}
H.~Shirani-Mehr, G.~Caire, and M.~J. Neely, ``{MIMO} downlink scheduling with
  non-perfect channel state knowledge,'' \emph{IEEE Trans. on Commun.},
  vol.~58, no.~7, pp. 2055--2066, Jul. 2010.

\bibitem{geng2013optimality}
C.~Geng, N.~Naderializadeh, A.~S. Avestimehr, and S.~A. Jafar, ``On the
  optimality of treating interference as noise,'' \emph{arXiv preprint
  arXiv:1305.4610}, 2013.

\bibitem{naderializadeh2013itlinq}
N.~Naderializadeh and A.~S. Avestimehr, ``{ITLinQ: A} new approach for spectrum
  sharing in device-to-device communication systems,'' \emph{IEEE Journal on
  Sel. Areas in Commun., to appear}, 2014.

\end{thebibliography}
